\documentclass[prl,twocolumn,showpacs,superscriptaddress,preprintnumbers]{revtex4-2}
\usepackage{amsmath,amssymb,amsfonts,float,graphics,epsfig,epstopdf,color,verbatim,tabularx,bm,multirow,appendix}

\usepackage{graphics}
\usepackage{xcolor}
\usepackage{dcolumn}
\usepackage{bm}
\usepackage{amsmath}
\usepackage{xspace}
\usepackage{placeins}
\usepackage{time}
\usepackage{orcidlink}
\usepackage{comment}
\usepackage[normalem]{ulem}
\pdfoutput=1
\usepackage{color}
\definecolor{LinkColor}{rgb}{0.256,0.439,0.588}
\usepackage{hyperref}
\hypersetup{
	colorlinks=true,
	citecolor=LinkColor,
	linkcolor=LinkColor,
	urlcolor=LinkColor
}

\newcommand{\beq}{\begin{equation}}
\newcommand{\eeq}{\end{equation}}
\newcommand{\beqn}{\begin{eqnarray}}
\newcommand{\eeqn}{\end{eqnarray}}

\newcommand{\re}{\textrm{e}}

\newcommand{\ve}[1]{\boldsymbol{#1}}

\newcommand\startsupplement{%
\newpage\clearpage
\setcounter{secnumdepth}{2}
\setcounter{table}{0}
\renewcommand{\thetable}{S\arabic{table}}
\setcounter{figure}{0}
\renewcommand{\thefigure}{S\arabic{figure}}
\setcounter{equation}{0}
\renewcommand{\theequation}{S\arabic{equation}}
\setcounter{section}{0}
\renewcommand{\thesection}{Section \Roman{section}}
\renewcommand{\thesubsection}{\Roman{section}. \Alph{subsection}}
}

\begin{document}

\title{Quantum Monte Carlo studies of U(1) lattice gauge models of Kondo breakdown }

\author{Gaopei Pan\orcidlink{0000-0003-3332-4557}}
\email{gaopei.pan@uni-wuerzburg.de}
\affiliation{Institut für Theoretische Physik und Astrophysik and Würzburg-Dresden Cluster of Excellence ct.qmat,
	Universität Würzburg, 97074 Würzburg, Germany}

\author{Fakher F. Assaad \orcidlink{0000-0002-3302-9243}}
\email{fakher.assaad@physik.uni-wuerzburg.de}
\affiliation{Institut für Theoretische Physik und Astrophysik and Würzburg-Dresden Cluster of Excellence ct.qmat,
	Universität Würzburg, 97074 Würzburg, Germany}

\date{\today}

\begin{abstract}
	In the local-moment regime, heavy fermions are most economically described by a compact U(1) gauge theory. Motivated by this perspective, we study a minimal compact U(1) lattice gauge model describing a spin chain coupled to two-dimensional Dirac conduction electrons.
	The spin chain is described by fermionic partons carrying spin and U(1) gauge charge. 
	The heavy-fermion quasiparticle is a bound state of a U(1) matter field carrying unit electric charge and U(1) gauge charge, and the fermionic parton. Using sign-problem-free determinant quantum Monte Carlo simulations, we identify two symmetry-equivalent regimes: a heavy-fermion metal with a sharp composite-fermion resonance and robust low-frequency transport, and a Kondo-breakdown metal with an incoherent resonance and vanishing low-frequency transport . For any finite lattice extent in the direction perpendicular to the chain, the Luttinger volume of the heavy-fermion phase counts both composite and conduction electrons, while in the Kondo-breakdown phase it counts only the conduction electrons. The evolution of the composite-fermion spectrum, dynamical spin structure factor, and optical conductivity provides a nonperturbative demonstration of gauge-mediated Kondo breakdown and establishes transport fingerprints of an orbital-selective Mott transition in the context of U(1) gauge theories of heavy fermions.
\end{abstract}

\maketitle
\noindent{\textcolor{blue}{\it Introduction}---}

The Kondo lattice model describes magnetic moments in a metallic environment~\cite{hewson1997kondo,coleman2001fermiliquids}. It plays an important role in a variety of domains, including heavy-fermion quantum criticality~\cite{coleman2001fermiliquids,si2001locally,si2013quantum,Mazza24}, bilayer systems of $^3$He~\cite{Neumann07,Beach2011}, the pseudogap phase of high-temperature superconductivity~\cite{zhang2020pseudogap}, twisted bilayer graphene~\cite{song2022magic,kang2104cascades,chou2023kondo,hu2023kondo,lau2025topological}, as well as magnetic adatoms on metallic 
surfaces~\cite{Toskovic16,danu2020kondo,danu2022spinchain,li2024artificially}. In this model, various phases and quantum phase transitions arise from the interplay between Kondo screening of the magnetic impurities, the magnetic Ruderman-Kittel-Kasuya-Yosida (RKKY) interaction~\cite{ruderman1954indirect,kasuya1956theory,yosida1957magnetic} between magnetic moments~\cite{doniach1977kondo}, and geometric frustration~\cite{burdin2002heavy}. The heavy-fermion phase of the Kondo lattice is a Fermi liquid with a Luttinger volume that counts both 
the conduction electrons and spin degrees of freedom. This defines the large Fermi surface~\cite{oshikawa2000,luttinger1960, ward1960}. As the effective mass grows, various instabilities of the heavy-fermion Fermi liquid can occur. In Hertz–Millis-type transitions, the heavy fermions in the vicinity of the large Fermi surface undergo magnetic ordering~\cite{coleman2001fermiliquids,si2013quantum}. Other instabilities involve the breakdown of Kondo screening and the concomitant reconstruction of the heavy-fermion Fermi surface~\cite{si2001locally,pepin2007kondo,de2005mott,komijani2019emergent,friedemann2009detaching,friedemann2010pnas,hartmann2010thermopower,Hofmann18}. This Kondo-breakdown transition is the focus of this work.

More broadly, Kondo-breakdown transitions may be viewed as a special limit of an orbital-selective Mott  transition~\cite{vojta2010orbital,de2008t,de2005mott,Dai2009}, where one sector (e.g., $f$ moments) localizes while the other remains itinerant.  This naturally requires a reconstruction  of the  Fermi-surface  and enables strange-metal transport without well-defined Landau quasiparticles~\cite{si2001locally,pepin2007kondo,si2014quantum,chen2023shot}. This perspective is particularly transparent in dimensionally mismatched systems, in which quasi-one-dimensional spin degrees of freedom are Kondo-coupled to higher-dimensional conduction electrons~\cite{danu2020kondo,danu2022spinchain,li2024artificially}, providing a tunable setting to access orbital-selective physics and its transport consequences. Importantly, local and mesoscopic probes can directly interrogate such non-quasiparticle transport: STM can resolve the spatially inhomogeneous Kondo hybridization and its collapse near criticality~\cite{danu2020kondo,danu2022spinchain,li2024artificially}, while recent shot-noise measurements in heavy-fermion nanowires reveal a strongly suppressed Fano factor that cannot be accounted for within a quasiparticle-based Fermi-liquid picture~\cite{chen2023shot,wang2024shot}, offering a direct window into strange-metal current carried by collective, strongly correlated excitations~\cite{li2023optical}.
\begin{figure}[!htbp]
	\includegraphics[width=\linewidth]{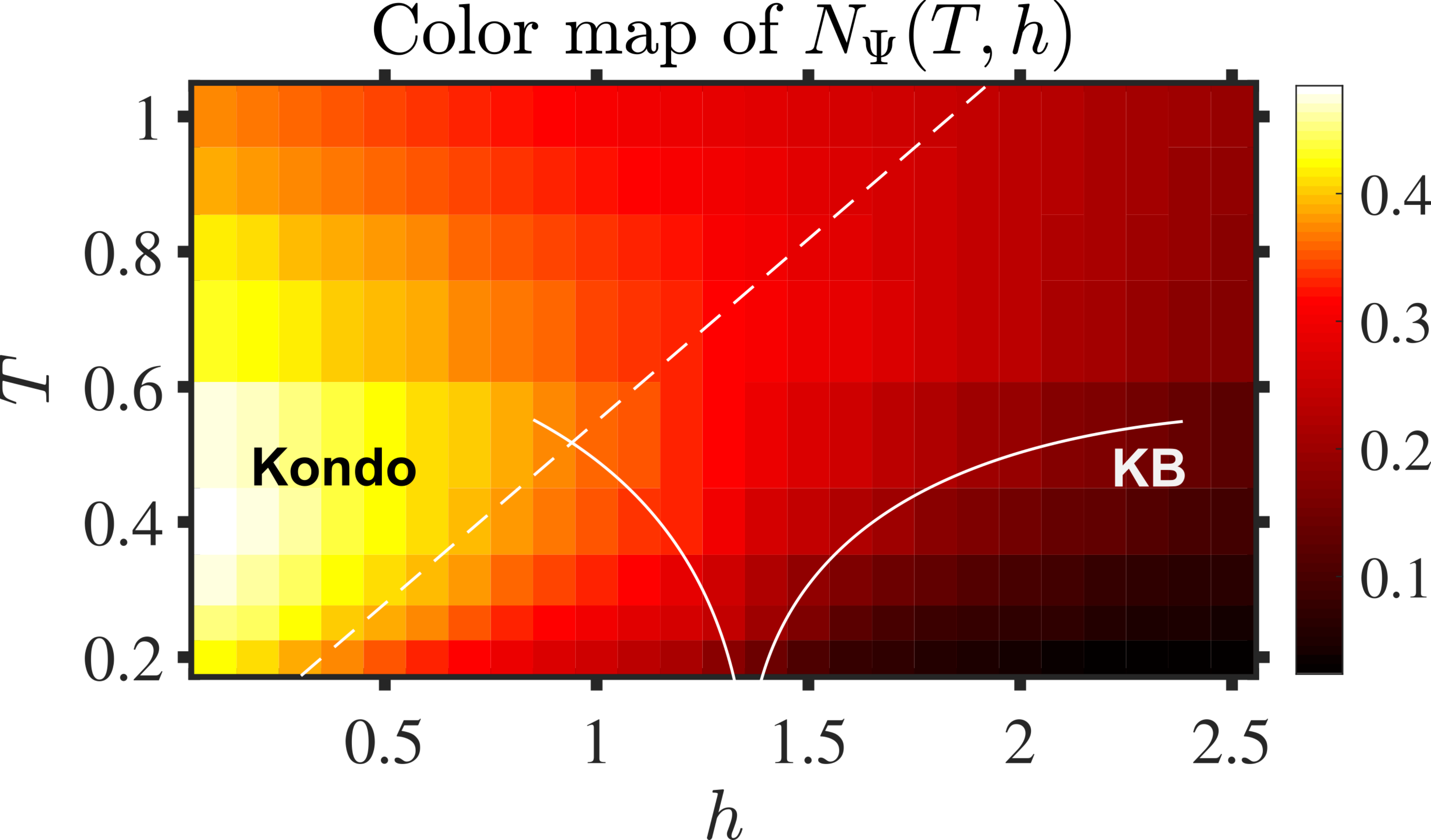}
	\caption{Color map of the composite-fermion zero-frequency density of states $N_{\Psi}(\omega = 0) $ 
		as a function of $h=h_m$, the strength of the electric-field fluctuations, and temperature $T$.   The dashed line traces the bare $h$ scale while the solid lines are guides to the eye for  crossover scales (see text).  
	}
	\label{fig:fig1}
\end{figure}

The pristine Kondo-breakdown transition does not involve symmetry breaking and has an odd number of impurity spins per unit cell~\cite{gazit2020fermi}. While the heavy-fermion Fermi 
liquid satisfies the Luttinger theorem~\cite{luttinger1960,ward1960}, the Kondo-breakdown phase violates it~\cite{PhysRevLett.98.026402,pepin2007kondo}. For dense systems, and owing to Oshikawa's topological proof of the Luttinger theorem, this violation necessitates topological degeneracy and is 
referred to as the FL$^*$ phase in which local moments and conduction electrons decouple~\cite{oshikawa2000,senthil2003fractionalized}. A model amenable to negative-sign-free quantum Monte Carlo simulations that exhibits a pristine Kondo-breakdown transition consists of a spin-1/2 chain antiferromagnetically coupled to two-dimensional Dirac electrons~\cite{danu2020kondo}. Here, the  vanishing density of states of the Dirac bath renders the Kondo coupling irrelevant at the  decoupled fixed point, thereby allowing for a stable Kondo-breakdown phase.
Assume that the size of the lattice hosting the Dirac electrons perpendicular to the chain is $L_y$ such that the unit cell hosts $L_y$ conduction electrons and a single spin-1/2 degree of freedom. In the Kondo-breakdown phase, the Luttinger volume, corresponding to the fraction of the Brillouin-zone volume $V_{BZ}$ enclosed by the Fermi surface, reads $\frac{V}{V_{BZ}} =\frac{1}{2}\,\mathrm{mod}(L_y,2)$, whereas in the heavy-fermion phase the spin participates in the Luttinger volume and the relation becomes $\frac{V}{V_{BZ}} =\tfrac{1}{2}\,\mathrm{mod}(L_y+1,2)$ 
~\cite{oshikawa2000,danu2020kondo}. 
Previous QMC simulations  established Kondo breakdown through spin and  single particle spectral properties~\cite{danu2020kondo,danu2022spinchain,li2024artificially,Mazza24}.  Transport has remained elusive  due to the difficulty computing current-current correlations of the composite fermion \cite{Danu21,PhysRevLett.98.026402,pepin2007kondo}.

Here we introduce a minimal compact and unconstrained U(1) lattice gauge model motivated by parton formulations of dimensionally mismatched Kondo-Heisenberg systems \cite{danu2020kondo}. 
Using determinant quantum Monte Carlo~\cite{blankenbecler1981monte,assaad2008world}, we identify Kondo-coherent and  Kondo-breakdown regimes from the composite-fermion spectrum, and we compute the optical conductivity along the chain by analytically continuing current--current correlations~\cite{Sandvik1,beach2004identifying,SHAO20231}. This provides controlled transport fingerprints of orbital selectivity across a pristine Kondo-breakdown transition. This direct access to composite-fermion transport is the main advance of the present work.

\noindent{\textcolor{blue}{\it Model and numerical method.}---}
Our model is motivated by the fermionic parton representation of the Kondo lattice model~\cite{danu2020kondo,danu2022spinchain,raczkowski2022breakdown,Praczkowski2024n,Danu21}. Here, the spin-$\tfrac{1}{2}$ degree of freedom is expressed as: 
$\ve{\hat{S}}_{r} = \frac{1}{2}\sum_{\sigma,\sigma'} \hat{f}_{r,\sigma}^\dagger \boldsymbol{\sigma}_{\sigma,\sigma'} \hat{f}_{r,\sigma'}$  
with constraint, $\sum_{\sigma} \hat{f}_{r,\sigma}^{\dagger} \hat{f}_{r,\sigma}^{\phantom\dagger} = 1$. In the above  $\ve{\sigma}$ is the vector of Pauli spin matrices.   The gauge degrees of freedom  correspond to the Hubbard-Stratonovich (HS) fields and the constraint endows  the fermionic parton with an emergent local U(1) gauge charge~\cite{read1983solution,auerbach1986kondo,saremi2007quantum,raczkowski2022breakdown}. 

\begin{equation}
	\begin{aligned}
		\hat{H}=& \sum_{\langle i , j\rangle, \sigma} t_{i, j} \ve{\hat{c}}_{i}^\dagger \ve{\hat{c}}_{j}+J \sum_r\left(\ve{\hat{f}}_{r}^{\dagger} \hat{U}_{r, r+a_x} \ve{\hat{f}}_{r+a_x} + \text{H.c.}\right) \\
		&+h \sum_b\hat{E}_b^2 + h_m \sum_r\hat{e}_r^2+V \sum_{r}\left(\ve{\hat{c}}_{r}^{\dagger} \hat{z}_r^{\dagger} \ve{\hat{f}}_{r}+ \text{H.c.} \right)
	\end{aligned}
	\label{eq:eq1}
\end{equation}
Here $\hat{U}_b=e^{i\hat{A}_b}$ and $\hat{z}_r=e^{i\hat{\phi}_r}$ are compact $U(1)$ variables defined on bonds and sites, respectively. The operator $\hat{E}_b$ is the electric field conjugate to $\hat{A}_b$, while $\hat{e}_r$ is the charge conjugate to $\hat{\phi}_r$; both have integer eigenvalues and satisfy $[\hat{E}_b,\hat{U}_{b'}]=\delta_{b,b'}\hat{U}_b$ and $[\hat{e}_r,\hat{z}_{r'}]=\delta_{r,r'}\hat{z}_r$. The terms proportional to $h$ and $h_m$ set the energy costs of nonzero $\hat{E}_b$ and $\hat{e}_r$, respectively, while $J$ and $V$ set the strengths of the $f$-fermion hopping and the local hybridization. Thus, $h$ controls the fluctuations of $\hat{U}_b$ and is not itself the microscopic Kondo coupling. 
We have suppressed the spin  indices for clarity. Since the U(1) gauge fields originate from a HS transformation, they are compact variables. The $c$-electrons reside on a two-dimensional square lattice with $\pi$ flux per plaquette, $t_{ij} = \pm t$, resulting in a Dirac semimetal with a vanishing density of states at the Fermi level.  As shown in the End Matter, total spin and total charge are  conserved:  Both the $c$-electrons and $f$-fermions carry spin while electric charge is carried by the $c$-electrons and the matter field $\hat{z}_r$.

$\hat{H}$ commutes with  $\hat{Q}_r =  \ve{\hat{f}}^{\dagger}_r \ve{\hat{f}}^{\phantom{\dagger}}_r   -  \hat{e}_r + \hat{E}_{r-a_x,r} - \hat{E}_{r,r+a_x}$   that generates local  U(1) gauge transformations.  With $\hat{T} = e^{i \sum_r \theta_r \hat{Q}_r}$   the fermions  and  gauge  fields transform as:
$		\hat{T}^{-1}\hat f_r \hat{T}=  e^{i\theta_r}\hat f_r,  $
$		\hat{T}^{-1}\hat z_r \hat{T}=  e^{i\theta_r}\hat z_r,  $
$		\hat{T}^{-1}\hat U_{r,r'} \hat{T}=  e^{i\theta_r}\hat U_{r,r'} e^{-i\theta_{r'}}, $
while the conduction electrons and electric fields  remain gauge neutral (i.e. commute with $\hat{T}$).  Each term in Eq.~\eqref{eq:eq1} is invariant under this local gauge transformation. Importantly, we do not impose  the Gauss law  constraint (a fixed value of $\hat{Q}_r$) on the Hilbert space,  but expect it to be  dynamically generated. In the large $h$ and $h_m$ limit, electric-field excitations are suppressed and the $f$-fermion parity is odd such that  $\hat{Q}_r=1$. In this limit, the model reduces to the Kondo--Heisenberg model with
$J_H=4J^2/h$ and $J_K=4V^2/h_m$ (see End Matter). For the parameters used here, with $h=h_m$, this limit corresponds to the Kondo-breakdown regime of Ref.~\cite{danu2020kondo}.

Central gauge-invariant objects in our analysis are the composite-fermion and bond variable
\begin{equation}
	\ve{\hat{\Psi}}_r = \hat{z}_r^{\dagger}\, \ve{\hat{f}}_r. \qquad
	\hat{W}_{r,r'}
	=
	\hat{z}_r^\dagger\hat{U}_{r,r'}\hat{z}_{r'}.
\end{equation}
$\ve{\hat{\Psi}}_r $ has no gauge charge but carries spin and electric
charge: it  has  the quantum numbers an electron.
In terms
of these variables, the hopping and hybridization terms become
$J\ve{\hat{\Psi}}_r^\dagger\hat{W}_{r,r'}\ve{\hat{\Psi}}_{r'}$
and
$V\ve{\hat{c}}_r^\dagger\ve{\hat{\Psi}}_r$, respectively.
This operator plays a crucial role in the low-temperature heavy-fermion physics of our model: it represents a naturally emergent quasiparticle that carries electric charge and spin but no gauge charge. 
If the $\ve{\hat{\Psi}}^{\dagger}_r$ operator creates a coherent quasiparticle that participates in the Luttinger volume,  
the ground state corresponds to the heavy-fermion phase. 
For $h,h_m\rightarrow0$,
$\hat{W}_{r,r'}$ becomes static, and the uniform choice
$W_{r,r'}=1$ gives the periodic Anderson model of conduction and
composite fermions (see End Matter). This does not require
$\hat{z}_r$ itself to have a nonzero expectation value.

Compared to more complicated quasiparticles in the full Kondo model \cite{Costi00,Raczkowski18,Danu21}, the composite fermion here takes a particularly simple form. In the DQMC simulations, $\hat{z}_r(\tau)$ is represented by a space and time dependent phase, such that including it does not introduce any additional complications for Wick’s theorem. This simplicity significantly improves the numerical efficiency and the quality of the composite-fermion Green’s function data. It also allows for new possibilities, namely the computation of  current-current correlations of the composite fermion.

We simulate Eq.~\eqref{eq:eq1} using the Algorithms for Lattice Fermions (ALF) \cite{ALF_v2.4}  implementation of sign-problem-free finite temperature DQMC \cite{blankenbecler1981monte,White89}  on lattices of size $L_x \times L_y$ for the $c$ electrons, with an $f$ chain of length $L_x$ embedded along the $x$ direction. Unless otherwise noted, we use inverse temperatures up to $\beta = 10$ and system sizes up to $L_x = L_y = 20$. Imaginary-time correlation functions are analytically continued to real frequencies using the ALF \cite{ALF_v2.4} implementation of  stochastic analytic continuation \cite{beach2004identifying,Sandvik1,SHAO20231}.
For simplicity, in the following we set $t = 1$, $V = 1$, and focus on the case $h = h_m$, which is controlled by a single parameter that tunes the strength of gauge and matter-field fluctuations.

\noindent{\textcolor{blue}{\it Results.}---} 

Gauge invariance forbids any expectation value of gauge-charged operators in accordance with Elitzur's theorem. Our analysis therefore focuses exclusively on gauge-invariant observables such as the composite fermion $\hat \Psi_r$ and spin correlations.

We begin by analyzing the local composite-fermion Green's function $G_{\Psi}(r,\tau) = \langle \hat{\Psi}_r(0)\,\hat{\Psi}_r^\dagger(\tau)\rangle$,
from which we approximate the zero-frequency density of states as
\begin{equation}
	N_{\Psi}(\omega=0) \simeq \beta\, G_{\Psi}(r=0,\tau=\beta/2).
\end{equation}
Fig.~\ref{fig:fig1} shows a color map of $N_{\Psi}(\omega = 0 )$ as a function
of the parameters $h$ and temperature $T$. In the End Matter section we show that  if we assume a Fermi-liquid form of the spectral function then $N_{\Psi}(\omega=0)  \propto \pi Z/v_F$   with $Z$ the quasiparticle residue and $v_F$ the Fermi velocity of  the composite fermion.   At our lowest temperatures, we observe that as a function of $h=h_m$, $N_{\Psi}(\omega=0)$ exhibits a sharp crossover interval $1.2\lesssim h\lesssim1.4$, which we associate with the transition between the Kondo-coherent and Kondo-breakdown regimes. The accessible system sizes and temperatures do not allow us to determine a precise critical coupling or critical exponents. 
Our  temperature range includes many scales. First, there is the bare $h$-scale. When the temperature scale  is above the $h$-scale, one can neglect the fluctuations of the gauge fields such that our  model effectively maps onto a non-interacting PAM, with coherence scale set by $V=1$ (see the End Matter). Hence, in the temperature range $ h < T < V $ we expect a coherent composite-fermion band with large density of states at the Fermi level.  As we cross the  $h$ scale in the  Kondo phase, $ h < h_c $, gauge fluctuations kick in  and  suppress the composite-fermion density of states.    The other  scales  correspond to crossover  scales, $T^{*} \propto |h -h_c|^{\nu z}$  with  $z$ ($\nu$)  the dynamical  (correlation length)  exponent.  The  white solid lines in Fig.~\ref{fig:fig1} 
are  guides to the eye for these crossover scales. 
In particular, on the Kondo side $T^{*}$ should correspond to the coherence scale of the composite fermion excitation, and on the Kondo-breakdown side to a localization scale of the composite fermions.
A  calculation of critical exponents is left for future work.

\begin{figure}[!htbp]
	\includegraphics[width=\linewidth]{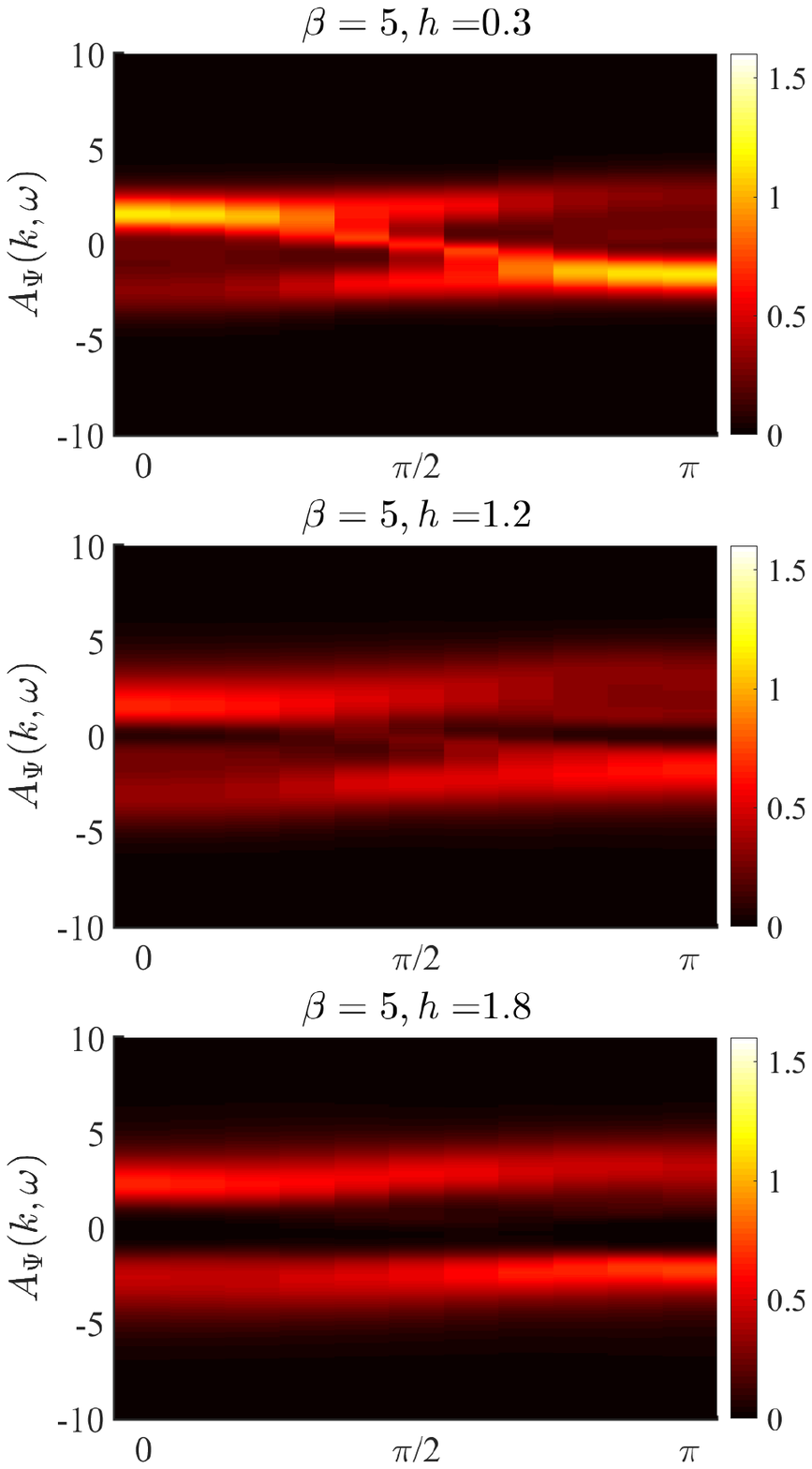}
	\caption{Composite-fermion spectral function $A_{\Psi}(k,\omega)$ at $\beta = 5$ for three representative values of the parameters $h$.
		For small $h$ (top panel), a clear hybridized low-energy band with substantial composite-fermion weight is visible,
		signaling the formation of heavy quasiparticles in the Kondo-coherent phase. Upon increasing $h$ (middle and bottom panels), the hybridization gradually disappears and the spectrum evolves into two separate, predominantly incoherent bands, indicating the destruction of the heavy quasiparticle band and the localization of the $f$ electrons characteristic of Kondo breakdown.}
	\label{fig:fig2}
\end{figure}

The composite-fermion spectral function $A_{\Psi}(k,\omega)$, where $\langle \hat{\Psi}_k(0) \hat{\Psi}_k^\dagger(\tau) \rangle= \int_{-\infty}^{\infty} \mathrm{d}\omega A_{\Psi}(k,\omega) \frac{\re^{-\tau \omega}}{\pi (1+\re^{-\beta \omega})}$, provides a more direct measure of Kondo hybridization between the conduction and localized electrons. In the small-$h$ phase, Fig.~\ref{fig:fig2} shows a clear hybridization and a coherent low-energy band with substantial composite-fermion weight. This feature signifies the emergence of heavy quasiparticles and corresponds to a Kondo-coherent state. Because the conduction electrons live on a lattice with finite transverse extent $L_y$, the conduction sector contains multiple discrete transverse momentum subbands. The additional dispersive features visible in Fig.~\ref{fig:fig2} originate from the hybridization between these finite-$L_y$ subbands and the composite fermion. As $h$ increases, the hybridization gradually disappears, leaving behind two separate incoherent bands.  The resulting spectrum indicates the destruction of the heavy quasiparticle band and the localization of the $f$ electrons—an unambiguous signature of Kondo breakdown.

\begin{figure}[!htbp]
	\centering
	\includegraphics[width=\linewidth]{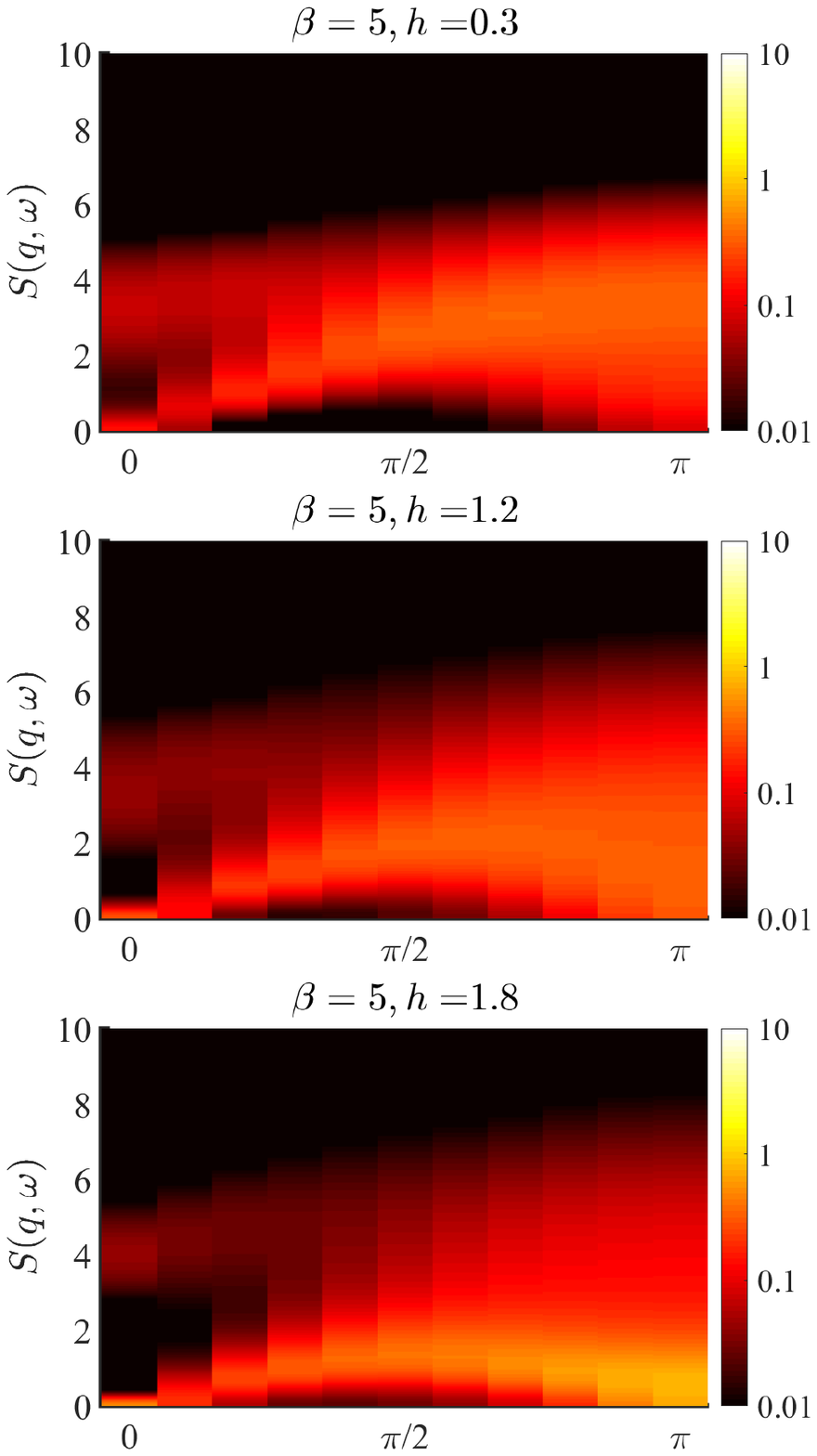}
	\caption{
		Dynamical spin structure factor of the $f$ chain, $S_f(q,\omega)$, at $\beta = 5$ for three values of the parameter $h$. For small $h$ (top panel), the spin response is broad and continuum-like, consistent with itinerant magnetic correlations mediated by hybridized quasiparticles in the Kondo-coherent phase. As $h$ increases (middle and bottom panels), a well-defined dispersive mode emerges and closely follows the spectrum of an isolated one-dimensional Heisenberg chain, demonstrating the recovery of local-moment behavior and corroborating the interpretation of $f$-electron Mott localization in the Kondo-breakdown phase.
	}
	\label{fig:fig3}
\end{figure}

To further characterize the two phases, we compute the dynamical spin structure factor of the $f$ chain, $S_f(q,\omega)$, shown in Fig.~\ref{fig:fig3}. 
Here $\langle \hat{S}_f(q,\tau) \hat{S}_f(-q,0) \rangle= \int_{0}^{\infty} \mathrm{d}\omega S_f(q,\omega) \frac{\re^{-\tau \omega}+\re^{-(\beta-\tau) \omega}}{\pi}$.  At small $h$, spin excitations are broad and continuum-like, reflecting itinerant magnetic correlations mediated by the hybridized quasiparticles. In contrast, at large $h$, the spin spectrum develops a well-defined dispersive mode, in good agreement with that of an isolated one-dimensional Heisenberg chain. This evolution demonstrates the recovery of local-moment behavior in the Kondo-breakdown phase and corroborates the interpretation of $f$-electron Mott localization. We note that the dynamics of the $f$-fermions at $V=0$ and in the limit of large values of $h$ is described by the  one-dimensional Heisenberg chain  (see supplemental material).

A key observable of our study is the transport along the one-dimensional $f$ chain. We evaluate the imaginary-time current--current correlation function $\Lambda(\tau) = \langle J_x(\tau) J_x(0)\rangle , $ and $ J_x = \sum_r \left( i \Psi_r \Psi_{r+a_x}^{\dagger} + \text{h.c.} \right)$,
and obtain the optical conductivity $\sigma'(\omega)$ along the chain by analytic continuation. The real part of the conductivity is related to $\Lambda(\tau)$ via
\begin{equation}
	\Lambda(\tau)
	= \int_0^{\infty} \frac{d\omega}{\pi}\,
	K(\tau,\omega)\,\sigma'(\omega),
	\label{eq:Lambda_sigma}
\end{equation}
with the kernel
\begin{equation}
	K(\tau,\omega)
	= \omega\,\frac{\cosh[\omega(\tau-\beta/2)]}{\sinh(\beta\omega/2)}.
	\label{eq:kernel}
\end{equation}

\begin{figure}[htbp]
	\centering
	\includegraphics[width=0.9\linewidth]{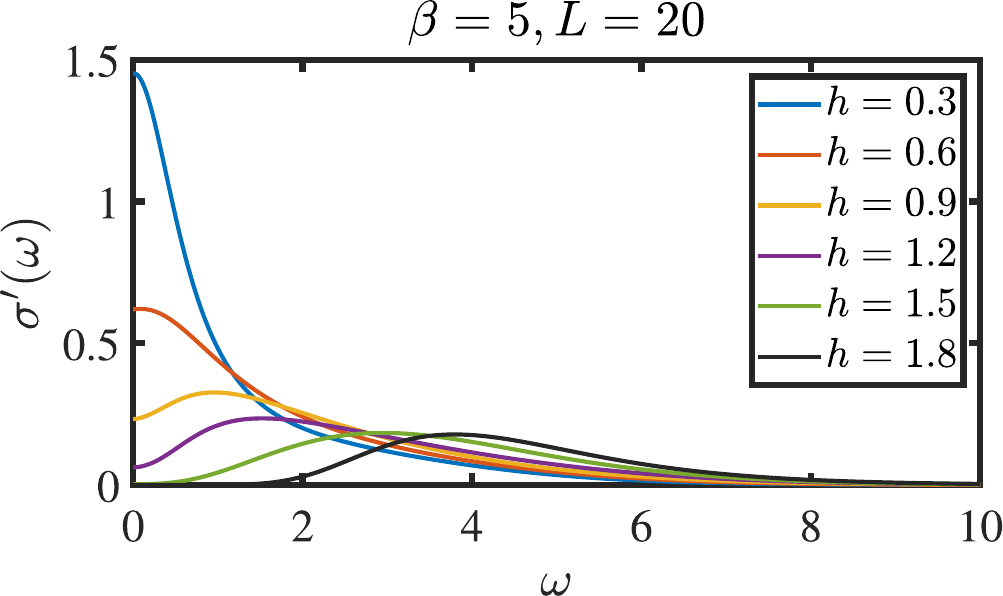}
	\caption{Real part of the composite-fermion optical response
		$\sigma'(\omega)$ at $\beta=5$. At small $h$, a pronounced
		low-frequency response is visible. With increasing $h$, low-frequency
		optical weight is depleted and transferred to a broad
		finite-frequency feature, providing a transport signature consistent
		with orbital-selective localization.
	}
	
	\label{fig:fig4}
\end{figure}

Figure~\ref{fig:fig4} shows the resulting $\sigma'(\omega)$ at fixed inverse temperature $\beta = 5$ for several values of the transverse field $h$. For small $h$ in the Kondo-coherent regime, the optical spectrum is dominated by a pronounced Drude-like peak at $\omega \approx 0$, indicating metallic transport carried by itinerant composite quasiparticles. Upon increasing $h$ and entering the Kondo-breakdown regime, the low-frequency weight and thus the Drude component are strongly suppressed, while spectral weight is transferred to a broad hump at finite frequencies. This evolution of $\sigma'(\omega)$ signals a metal-to-Mott-insulator transition in the $f$ sector, whereas the Dirac $c$ electrons remain itinerant, providing direct dynamical evidence for an orbital-selective Mott transition.

\noindent{\textcolor{blue}{\it Summary and conclusions.}---}
The one-dimensional spin-1/2 chain antiferromagnetically coupled to two-dimensional Dirac conduction electrons is a pristine example of 
a Kondo breakdown transition, defined by the absence of symmetry breaking between the two phases and a change of the Luttinger volume across the transition accounting 
for the localization of the spin degrees of freedom~\cite{danu2020kondo}. In this work, we introduce a minimal lattice-gauge model that captures this 
physics and allows us to go beyond previous DQMC studies that focused on the Kondo lattice model. In particular, the lattice-gauge formulation allows for the
calculation of transport properties that capture the defining signatures of the Kondo breakdown transition. In other words, our calculations 
provide transport signatures of an orbital-selective Mott transition in a controlled lattice simulation.

The quantum phase transition is driven by a coherence-incoherence transition of the composite fermion in which the particle nature of the composite fermion is lost.  This is seen both in the transport properties and in the spectral function. These observations are in very good agreement with recent shot-noise experiments that signal the loss of particle-like transport, where a strongly suppressed Fano factor signals that the current is no longer carried by long-lived particle-like excitations \cite{wang2024shot,chen2023shot}. The breakdown of coherence of the composite fermion shifts spectral weight from low to high frequencies. This phenomenon leads to an enhancement of the  quantum Fisher information as  recently observed in the context of the Kondo-breakdown transition \cite{Mazza24}

Our approach is very similar to the Fermi-surface reconstruction without symmetry breaking introduced in Ref.~\cite{gazit2020fermi} in the framework of a $\mathbb{Z}_2$ gauge theory with orthogonal fermions 
\cite{Nandkishore12,Hohenadler19} and matter fields. 
Looking forward, 
our compact U(1) lattice gauge description of Kondo-breakdown physics opens many new avenues. It is known that lattice gauge theories can be efficiently 
simulated using Hybrid Monte Carlo (HMC) algorithms~\cite{Duane87,Beyl17}, especially in the deconfined phase of compact U(1) gauge theories~\cite{Feng25}. 
Achieving larger systems with the lattice-gauge formulation will allow us to study critical behavior not only of Kondo breakdown transitions, but also magnetic order--disorder transitions in dimensionally mismatched Kondo lattices~\cite{danu2022spinchain,LiuZH22a,Frank22}.

{\it{Acknowledgment.-}} The authors thank Zi Hong Liu, Jo\~ao C. In\'acio,  B. Danu and S. Biswas for discussions. GPP acknowledges the  German Research Foundation (DFG)  through the W\"urzburg-Dresden Cluster of Excellence on Complexity and Topology in Quantum Matter -- \textit{ct.qmat} (EXC 2147, Project No.\ 390858490) as well as through the 
AS 120/16-2 (Project number 493886309) that is part of the collaborative research project SFB QMS funded by the Austrian Science Fund (FWF) F 86.
FFA acknowledges support from the DFG under the grant DA 2805/2 (Project number 528834426).
FFA  and  GPP  gratefully acknowledge the Gauss Centre for Supercomputing e.V.\ (www.gauss-centre.eu) for funding this project by providing computing time on the GCS Supercomputer SUPERMUC-NG at Leibniz Supercomputing Centre (www.lrz.de),   (project number pn73xu)     
as  well  as  the scientific support and HPC resources provided by the Erlangen National High Performance Computing Center (NHR@FAU) of the Friedrich-Alexander-Universit\"at Erlangen-N\"urnberg (FAU) under the NHR project b133ae. NHR funding is provided by federal and Bavarian state authorities. NHR@FAU hardware is partially funded by the German Research Foundation (DFG) -- 440719683. 
Calculations  were  carried out  with the  ALF-package~\cite{ALF_v2.4}

\bibliography{ref.bib}
\bibliographystyle{apsrev4-2}

\newpage
\newpage

\begin{widetext}
	
	\section{End Matter}
	
	\section{ Symmetries}
	\label{sec:Kondo_Heisenberg_limit}
	In this section, we show that   in the large $h$ and large $h_m$ limits, our model maps onto the Kondo-Heisenberg model considered in Ref.~\cite{danu2020kondo}. 
	Our  Hamiltonian reads:
	\begin{equation}
		\begin{aligned}
			H= \sum_{\langle i , j\rangle, \sigma} t_{i, j} \ve{\hat{c}}_{i}^\dagger \ve{\hat{c}}_{j}+J \sum_r\left(\ve{\hat{f}}_{r}^{\dagger} \hat{U}_{r, r+a_x} \ve{\hat{f}}_{r+a_x} + \text{H.c.}\right) 
			+h \sum_b\hat{E}_b^2 + h_m \sum_r\hat{e}_r^2+V \sum_{r}\left(\ve{\hat{c}}_{r}^{\dagger} \hat{z}_r^{\dagger} \ve{\hat{f}}_{r}+ \text{H.c.} \right),
		\end{aligned}
		\label{eq:Ham_SM}
	\end{equation}
	where the link and  electric  field  operators satisfy the commutation relations: 
	\begin{equation}
		\left[\hat{E}_b, \hat{U}_{b'}\right] = \delta_{b,b'} \hat{U}_b, \qquad
		\left[\hat{e}_r, \hat{z}_{r'}\right] = \delta_{r,r'} \hat{z}_r.
	\end{equation}
	The  model corresponds to a so-called unconstrained  lattice  gauge  theory.  By this, we mean that there is a locally conserved  quantity, 
	\begin{equation}
		\hat{Q}_r =  \ve{\hat{f}}^{\dagger}_r \ve{\hat{f}}^{\phantom{\dagger}}_r   -  \hat{e}_r + \hat{E}_{r-a_x,r} - \hat{E}_{r,r+a_x}  \label{eq:Gauss_law}
		\quad \text{with} \quad [\hat{Q}_r, H] = 0.
	\end{equation}
	This corresponds to the Gauss law. 
	$\hat{Q}_{r}$ is the  generator of local gauge transformations:   $\hat{T} = e^{i \sum_r \theta_r \hat{Q}_r}$ 
	Thereby: 
	$\hat{T}^{-1} \hat{f}_r \hat{T} = e^{i \theta_r}  \hat{f}_r$, 
	$ \hat{T}^{-1} \hat{U}_{r,r+a_x} \hat{T}  = e^{i \theta_r} \hat{U}_{r,r+a_x} e^{-i \theta_{r+a_x}}$, and 
	$ \hat{T}^{-1} \hat{z}_r \hat{T} = e^{i \theta_r} \hat{z}_r$.  
	
	Here ``unconstrained'' means that no fixed eigenvalue of
	$\hat{Q}_r$ is imposed on the Hilbert space. Since
	$[\hat{Q}_r,\hat{H}]=0$, the Hilbert space decomposes into
	superselection sectors labeled by the eigenvalues of $\hat{Q}_r$,
	and the local $U(1)$ symmetry is exact in every sector. In the
	$\hat{Q}_r=1$ sector, the large-$h,h_m$ limit suppresses
	$\hat{E}_b\neq0$ and $\hat{e}_r\neq0$, thereby recovering the
	single-occupancy condition
	$\ve{\hat{f}}_r^\dagger\ve{\hat{f}}_r=1$ of the standard parton
	representation.
	
	Aside from the local U(1) gauge symmetry, our  model also enjoys global charge conservation.  The generator of this symmetry is given by: 
	\begin{equation}
		\hat{N} = \hat{N}
		=
		\sum_i
		\ve{\hat{c}}_i^\dagger
		\ve{\hat{c}}_i
		+
		\sum_r\hat{e}_r, \qquad \text{with} \qquad [\hat{N}, H] = 0.
	\end{equation}
	
	As a consequence, the $\hat{z}_r$ field  carries  both  gauge  and electric charge, whereas
	$\ve{\hat{\Psi}}_r=\hat{z}_r^\dagger\ve{\hat{f}}_r$ has the same
	physical charge as the conduction electron and is gauge neutral.
	
	Another symmetry of the model  is  the  particle-hole symmetry $\hat{P}$ defined as:   
	\begin{equation}
		\hat{P}^{-1} \hat{c}^{\dagger}_{i,\sigma} \hat{P} = (-1)^i \hat{c}_{i,\sigma}, \qquad
		\hat{P}^{-1} \hat{f}^{\dagger}_{r,\sigma} \hat{P} = -(-1)^i \hat{f}_{r,\sigma}, \qquad
		\hat{P}^{-1} \hat{z}^{\dagger}_{r}  \hat{P} =    \hat{z}_r, \qquad
		\hat{P}^{-1} \hat{U}^{\dagger}_{b}  \hat{P} =    \hat{U}_b.
	\end{equation}
	As a consequence of this  symmetry,  $ \langle \ve{\hat{f}}^{\dagger}_r \ve{\hat{f}}^{\phantom{\dagger}}_r \rangle = 1 $.
	Finally, the  total spin  is  conserved,
	\begin{equation}
		\ve{\hat{S}}^{\mathrm{tot}}
		=
		\sum_i
		\ve{\hat{c}}_i^\dagger
		\frac{\boldsymbol{\sigma}}{2}
		\ve{\hat{c}}_i
		+
		\sum_r
		\ve{\hat{f}}_r^\dagger
		\frac{\boldsymbol{\sigma}}{2}
		\ve{\hat{f}}_r \qquad \text{with} \qquad [\ve{S}^{tot}, H] = 0,
	\end{equation}
	such that the model has global SU(2) symmetry. As a consequence of the above, both $f$- and 
	$c$-fermions carry spin-$\tfrac{1}{2}$ quantum numbers.
	
	\section{ Strong-coupling limit:  mapping to the Kondo-Heisenberg model  and Kondo breakdown}
	In the large $h$ and  $h_m$ limits  the electric fields  are expensive,  such that in this limit $\hat{Q}_r = 
	\ve{\hat{f}}^{\dagger}_r \ve{\hat{f}}^{\phantom{\dagger}}_r$.   Since fluctuations of the density 
	of $f$-fermions on a site necessitate the excitation of quanta of electric field, we have $\ve{\hat{f}}^{\dagger}_r \ve{\hat{f}}^{\phantom{\dagger}}_r = 1 $ in this limit.   
	We can now carry out a perturbative expansion in $1/h$ and $1/h_m$ around this limiting case to obtain, in second order, the effective Kondo-Heisenberg model.  As mentioned above, hopping of an $f$-fermion from site $r$ to its nearest-neighbor site $r'$ excites an electric-field quantum  with an energy cost of order $h$. The only way to remove this excitation is for the $f$-fermion to  hop back from $r'$ to $r$. These virtual processes generate the effective  $\mathrm{SU}(2)$ Kondo-Heisenberg model:
	\begin{equation}
		\hat{H}_{eff} =  \sum_{\langle i , j\rangle, \sigma} t_{i, j} \ve{\hat{c}}_{i}^\dagger \ve{\hat{c}}_{j}  
		-\frac{V^2}{h_m} \sum_{r} \left( \hat V_{r}^{\dagger} \hat V_{r} + \hat V_{r} \hat V_{r}^{\dagger} \right)
		-\frac{J^2}{h} \sum_{\langle r r' \rangle} \left( \hat D_{rr'}^{\dagger} \hat D_{rr'} + \hat D_{rr'} \hat D_{rr'}^{\dagger} \right),
	\end{equation}
	with 
	$\hat D_{rr'} = \ve{\hat{f}}_{r}^{\dagger} \ve{\hat{f}}_{r'}$  and $\hat V_{r} = \ve{\hat{f}}_{r}^{\dagger} \ve{\hat{c}}_{r}$.  
	The equivalence of the two models  is made explicit by using the equation 
	\begin{equation}
		\hat D_{rr'}^{\dagger} \hat D_{rr'} + \hat D_{rr'} \hat D_{rr'}^{\dagger}	 = - 4 \hat{\mathbf{S}}_{r} \cdot \hat{\mathbf{S}}_{r'} + 1
	\end{equation}
	valid on the Hilbert space with one $f$-fermion per site, and the analogous equation for the Kondo term: 
	$\hat V_{r}^{\dagger} \hat V_{r} + \hat V_{r} \hat V_{r}^{\dagger}	 = - 4 \hat{\mathbf{S}}_{r} \cdot \hat{\mathbf{s}}_{r} + 1 $.
	Here, 
	$\ve{S}_r =  \frac{1}{2} \ve{\hat{f}}_{r}^{\dagger} \boldsymbol{\sigma} \ve{\hat{f}}_{r}$  is the spin operator of the $f$-electron.  Hence,
	\begin{equation}
		\label{eq:kondo_heisenberg}
		\hat{H}_{eff} =  \sum_{\langle i , j\rangle, \sigma} t_{i, j} \ve{\hat{c}}_{i}^\dagger \ve{\hat{c}}_{j}  
		+ J_K \sum_{r} \hat{\mathbf{S}}_r \cdot \hat{\mathbf{s}}_r 
		+ J_H \sum_{\langle r r' \rangle}  \hat{\mathbf{S}}_r \cdot \hat{\mathbf{S}}_{r'} + \text{const.}
	\end{equation}	
	Here,
	\begin{equation}
		\label{eq:effective_exchanges}
		J_K=\frac{4V^2}{h_m},
		\qquad
		J_H=\frac{4J^2}{h},
	\end{equation}
	and $\ve{s}_r =  \frac{1}{2} \ve{\hat{c}}_{r}^{\dagger} \boldsymbol{\sigma} \ve{\hat{c}}_{r}$ is the spin operator 
	of the $c$-electrons.
	Provided that $J_K = 4 V^2/h_m$ is small, a power-counting argument presented in Ref.~\cite{danu2020kondo}  shows that this interaction is irrelevant at the decoupled fixed point, thereby describing a Kondo-breakdown phase.
	
	\section{ Weak-coupling limit: the heavy-fermion phase.}
	
	We now show that in the weak-coupling limit,  the model maps onto the large-$N$ mean-field theory of Eq.~\ref{eq:kondo_heisenberg}, thereby showing that this limit corresponds to the heavy-fermion phase. Let us  recall that the composite-fermion operator and bond operator
	are  defined as:
	\begin{equation}
		\ve{\hat{\Psi}}_r
		=
		\hat{z}_r^\dagger\ve{\hat{f}}_r,
		\qquad
		\hat{W}_{r,r+a_x}
		=
		\hat{z}_r^\dagger
		\hat{U}_{r,r+a_x}
		\hat{z}_{r+a_x}.
		\label{eq:gauge_invariant_weak_limit}
	\end{equation}
	Both operators commute with all local gauge generators $\hat{Q}_{r'}$.
	Since in our  representation $\hat{z}_r \hat{z}^{\dagger}_r = 1$, we  can rewrite  the  Hamiltonian as: 
	\begin{equation}
		\begin{aligned}
			\hat{H}={}
			\sum_{\langle i,j\rangle,\sigma}
			t_{i,j}
			\ve{\hat{c}}_i^\dagger
			\ve{\hat{c}}_j
			+
			J\sum_r
			\left(
			\ve{\hat{\Psi}}_r^\dagger
			\hat{W}_{r,r+a_x}
			\ve{\hat{\Psi}}_{r+a_x}
			+\text{H.c.}
			\right)
			+
			h\sum_b\hat{E}_b^2
			+
			h_m\sum_r\hat{e}_r^2
			+
			V\sum_r
			\left(
			\ve{\hat{c}}_r^\dagger
			\ve{\hat{\Psi}}_r
			+\text{H.c.}
			\right).
			\label{eq:gauge_invariant_H}
		\end{aligned}
	\end{equation}
	
	As $h,h_m\rightarrow0$, temporal fluctuations of the compact fields
	are suppressed and $\hat{W}_{r,r+a_x}$ becomes static. The
	translationally invariant zero-flux sector has
	$W_{r,r+a_x}=1$, for which Eq.~\eqref{eq:gauge_invariant_H} reduces to
	\begin{equation}
		\hat{H}_{\mathrm{PAM}}={}
		\sum_{\langle i,j\rangle,\sigma}
		t_{i,j}
		\ve{\hat{c}}_i^\dagger
		\ve{\hat{c}}_j
		+
		J\sum_r
		\left(
		\ve{\hat{\Psi}}_r^\dagger
		\ve{\hat{\Psi}}_{r+a_x}
		+\text{H.c.}
		\right)
		+
		V\sum_r
		\left(
		\ve{\hat{c}}_r^\dagger
		\ve{\hat{\Psi}}_r
		+\text{H.c.}
		\right).
		\label{eq:periodic_anderson_limit}
	\end{equation}
	This is the periodic-Anderson form corresponding to the large-$N$
	heavy-fermion saddle point of the Kondo--Heisenberg model,
	Eq.~\eqref{eq:kondo_heisenberg}. Importantly, this reduction involves
	only the gauge-invariant fields $\ve{\hat{\Psi}}_r$ and
	$\hat{W}_{r,r+a_x}$ and does not require
	$\langle\hat{z}_r\rangle\neq0$ or
	$\langle\hat{U}_{r,r+a_x}\rangle\neq0$.
	
	\section{A proxy for the quasiparticle  weight of the composite fermion. }
	Consider the quantity: 
	\begin{equation}  
		Z_{\text{eff}} = \beta  G_{\Psi}(0,\tau=\beta/2)  = \frac{1}{L} \sum_k  \int d \omega g_\beta(\omega) A_{\Psi}(k,\omega) \quad \text{with} \quad g_\beta(\omega) = \beta g(\beta\omega), \quad g(\omega)=\frac{1}{\pi} \frac{e^{- \omega/2}}{1 +  e^{-\omega}}
	\end{equation}
	Since $\int d\omega g(\omega) = 1$,  $\lim_{\beta \to \infty} g_\beta(\omega) = \delta(\omega)$. At  finite temperatures, we can approximate  $g_\beta(\omega)$  by a box function of height     $\beta$, and  width  $1/\beta$ centered at $\omega = 0$. 
	Let  us now assume  that 
	\begin{equation}
		A_{\Psi}(k,\omega) = Z \delta(v_F  k  -  \omega ) + A^{inc}_{\Psi}(k,\omega), 
	\end{equation}
	where  $A^{inc}_{\Psi}(k,\omega)$ is the incoherent part of the spectral function. 
	In the heavy-fermion phase, the  quasiparticle weight $Z$ is finite and  $A^{inc}_{\Psi}(k,\omega)$ is negligible at low  energies.  In this case,  we have $Z_{\text{eff}}   = \pi Z /v_F$.  In the Kondo-breakdown phase, $Z = 0$ and $A^{inc}_{\Psi}(k,\omega)$ is negligible at low energies.  In this case, $Z_{\text{eff}} \simeq 0$.  Hence, $Z_{\text{eff}}$ serves as a proxy for the quasiparticle weight of the composite fermion.
	
\end{widetext}

\startsupplement

\begin{widetext}
	
	
	\begin{center}
		{\bf \uppercase{Supplemental Materials for \\[0.5em]
				Quantum Monte Carlo studies of U(1) lattice gauge models of Kondo breakdown}}
	\end{center}
	
	\section{Relation to the Kondo-Heisenberg:  field theory approach.}
	
	Aside from the exact mapping onto the Kondo-Heisenberg Hamiltonian in the large $h$ and $h_m$ limits, we can also motivate the effective compact U(1) lattice gauge model used in the main text from a field-theory perspective \cite{saremi2007quantum,raczkowski2022breakdown}. We start from the Kondo-Heisenberg model
	\begin{equation}
		\label{eq:kondo_heisenberg}
		\hat{H}
		=  \sum_{\langle i,j\rangle,\sigma} \left( \hat{c}^{\dagger}_{i} t_{i,j}\hat{c}_{j} + \text{h.c.} \right)
		+ \frac{J_K}{2} \sum_{r} \hat{c}^{\dagger}_{r} \boldsymbol{\sigma} \hat{c}_{r} \cdot \hat{\mathbf{S}}_{r}
		+ J_H \sum_{\langle r,r'\rangle} \hat{\mathbf{S}}_{r} \cdot \hat{\mathbf{S}}_{r'}.
	\end{equation}
	We adopt an  Abrikosov  fermion representation of the spin-1/2 degree of freedom:
	\begin{equation}
		\hat{\mathbf{S}}_{r} = \frac12 \hat{f}^{\dagger}_{r} \boldsymbol{\sigma} \hat{f}_{r}, \qquad
		\hat{f}^{\dagger}_{r} = \bigl(\hat{f}^{\dagger}_{r,\uparrow},\hat{f}^{\dagger}_{r,\downarrow}\bigr),
	\end{equation}
	Introducing the fermion bilinears
	\begin{equation}
		\hat{V}_{r} = \hat{f}^{\dagger}_{r}\hat{c}_{r}, \qquad
		\hat{D}_{b} = \hat{f}^{\dagger}_{r}\hat{f}_{r'} \quad (b=\langle r,r'\rangle),
	\end{equation}
	the Kondo and Heisenberg terms can be written as perfect squares,
	\begin{equation}
		\hat{H}
		= -t \sum_{\langle i,j\rangle} (\hat{c}^{\dagger}_{i}\hat{c}_{j} + \text{h.c.})
		- \frac{J_k}{8} \sum_{r} \Bigl[(\hat{V}_{r}+\hat{V}^{\dagger}_{r})^{2} + (i\hat{V}_{r}-i\hat{V}^{\dagger}_{r})^{2}\Bigr]
		- \frac{J_h}{8} \sum_{b=\langle r,r'\rangle} \Bigl[(\hat{D}_{b}+\hat{D}^{\dagger}_{b})^{2} + (i\hat{D}_{b}-i\hat{D}^{\dagger}_{b})^{2}\Bigr]
		+ \hat{H}_U .
	\end{equation}
	Here
		$\hat{H}_U=\frac{U}{2}\sum_r
		(\hat{n}_r^f-1)^2$, with
		$\hat{n}_r^f=\hat{f}_r^\dagger\hat{f}_r$,
		enforces the single-occupancy condition in the limit
		$U\rightarrow\infty$.
	Decoupling these squares with complex Hubbard-Stratonovich fields $b_{r}$ (on sites) and $\chi_{b}$ (on bonds), and taking the limit $U\to\infty$, the partition function can be written as a path integral
	\begin{equation}
		Z = \mathrm{Tr}\,e^{-\beta\hat{H}}
		= \int \mathcal{D}\{f^{\dagger}f\}\,\mathcal{D}\{c^{\dagger}c\}\,\mathcal{D}\{\chi_{b}\}\,\mathcal{D}\{b_{r}\}\,\mathcal{D}\{a_{0}\}\, e^{-S} ,
	\end{equation}
	with an action of the form
	\begin{align}
		S = \int_{0}^{\beta} \! d\tau \Biggl\{ &
		\frac{2}{J_h}\sum_{b} |\chi_{b}(\tau)|^{2}
		+ \frac{2}{J_k}\sum_{r} |b_{r}(\tau)|^{2}
		+ \sum_{i,j} c^{\dagger}_{i}(\tau)\,[\partial_{\tau}\delta_{ij}-T_{ij}]\,c_{j}(\tau) \notag \\
		&+ \sum_{r} |b_{r}(\tau)|\Bigl[e^{i\varphi(r,\tau)} f^{\dagger}_{r}(\tau)c_{r}(\tau)+\text{h.c.}\Bigr]
		+ \sum_{r} f^{\dagger}_{r}(\tau)[\partial_{\tau}+i a_{0}(r,\tau)]f_{r}(\tau) \notag\\
		&+ \sum_{b=\langle r,r'\rangle} |\chi_{b}(\tau)|\Bigl[f^{\dagger}_{r}(\tau)e^{i\int_{r}^{r'}\mathbf{a}(\ell,\tau)\cdot d\ell}\,f_{r'}(\tau)+\text{h.c.}\Bigr] \Biggr\},
	\end{align}
	where we have parametrized $b_{r}(\tau)=|b_{r}(\tau)|e^{i\varphi(r,\tau)}$ and $\chi_{b}=|\chi_{b}|e^{i\int_{r}^{r'}\mathbf{a}\cdot d\ell}$ for $b = (r,r')$.  The phases $\varphi(r,\tau)$ and the link variables $e^{i\int_{r}^{r'}\mathbf{a}\cdot d\ell}$, together with the temporal component $a_{0}(r,\tau)$, transform as a compact U(1) gauge field under
	$f_{r}(\tau)\!\to e^{i\theta_{r}(\tau)}f_{r}(\tau)$, while the conduction electrons $c_{r}$ remain gauge neutral.  
	In particular,  the local gauge transformation
	\begin{align}
		\ve{a}(\ell,\tau) &\to \ve{a}(\ell,\tau) - \nabla \theta(\ell,\tau),    \nonumber \\
		a_{0}(\ell,\tau) &\to a_{0}(\ell,\tau) - \partial_{\tau} \theta(\ell,\tau), \qquad \text{ and } \nonumber \\ 
		\varphi(r,\tau) &\to \varphi(r,\tau) + \theta(r,\tau), 
	\end{align}
	can be absorbed in  the canonical transformation of the $f$-fermions:  $f_{r}(\tau)\to e^{i\theta(r,\tau)}f_{r}(\tau)$.
	
	To see how  this relates  to our gauge theory  we will have to take into account  the  Gauss law  described in the End Matter. In particular, we will assume, as is the case in the strong-coupling limit, that the Gauss law is dynamically imposed  and,  
	again, as in the strong-coupling limit, takes the value $\hat{Q}(r) = 1$.  The  Hamiltonian we will consider to derive the appropriate action is  then 
	\begin{equation}
		\begin{aligned}
			\hat{H}_{U(1)}= \sum_{ i , j} T_{i, j} \ve{\hat{c}}_{i}^\dagger \ve{\hat{c}}_{j}+J \sum_r\left(\ve{\hat{f}}_{r}^{\dagger} \hat{U}_{r, r+a_x} \ve{\hat{f}}_{r+a_x} + \text{H.c.}\right) 
			+h \sum_b\hat{E}_b^2 + h_m \sum_r\hat{e}_r^2+V \sum_{r}\left(\ve{\hat{c}}_{r}^{\dagger} \hat{z}_r^{\dagger} \ve{\hat{f}}_{r}+ \text{H.c.} \right) + \lambda \sum_r \left(\hat{Q}_r-1\right)^2
		\end{aligned}
	\end{equation}
	In the above we have extended the Hamiltonian  with a term that is symmetry-allowed and  that  partially imposes the Gauss law.  
	We can now carry out  a path integral of the $U(1)$ gauge  theory.  For simplicity, we use a non-compact formulation and adopt the representation: 
	\begin{equation}
		\hat{U}_b = e^{iA_b}, \qquad \hat{E}_b =  \frac{\partial}{\partial i A_b},  \qquad  
		\hat{z}_r = e^{i\varphi_r}, \qquad \text{and}, \qquad \hat{e}_r =  \frac{\partial}{\partial i \varphi_r}.
	\end{equation}
	A  standard calculation gives: 
	\begin{align}
		S_{U(1)} = \int_{0}^{\beta} \! d\tau \Biggl\{ &
		\frac{1}{4 \lambda}\sum_{r} |a_{0,r}(\tau)|^{2}
		+ \sum_{i,j} c^{\dagger}_{i}(\tau)\,[\partial_{\tau}\delta_{ij}-T_{ij}]\,c_{j}(\tau) \notag \\
		&+ \sum_{r} \Bigl[e^{i\varphi_{r}(\tau)} f^{\dagger}_{r}(\tau)c_{r}(\tau)+\text{h.c.}\Bigr]
		+ \sum_{r} f^{\dagger}_{r}(\tau)[\partial_{\tau}+i a_{0,r}(\tau)]f_{r}(\tau) 
		+ \sum_{\langle r,r'\rangle} \Bigl[f^{\dagger}_{r}(\tau)e^{i\int_{r}^{r'}\mathbf{a}(\ell,\tau)\cdot d\ell}\,f_{r'}(\tau)+\text{h.c.}\Bigr]  \notag \\ 
		& + \frac{1}{h} \sum_{\langle r,r'\rangle} \left(  \int_{r}^{r'} \dot{\mathbf{a}}(\ell,\tau)\cdot d\ell    +  a_{0,r}(\tau) - a_{0,r'}(\tau)\right)^2  + \frac{1}{h_m} \sum_{r} \left( \dot{\theta}(r,\tau) + a_{0,r}(\tau) \right)^2  - i \sum_{r} a_{0,r}(\tau) \Biggr\}.
	\end{align}	
	Here,  we have  denoted by  $a_0(r,\tau)$  the  field that we introduce to impose the Gauss law.
	The  similarities and differences between the two actions are clearly  visible, and  we note the following. 
	\begin{itemize}
		\item  In the strong-coupling limit, $h,h_m \rightarrow \infty$, the  last two  terms in  $S_{U(1)}$ are suppressed. As  shown in the  End Matter,  the   Gauss law  is  exactly imposed in this limit,  such that $\lambda \rightarrow \infty$.    Hence, aside  from  amplitude fluctuations of the bosonic fields, both actions are equivalent. In this limit, both actions enjoy local $U(1)$ symmetry.  
		\item  At finite values of $h$ and $h_m$,   we expect a finite gap  between  different Gauss law sectors.  As a consequence,    $\lambda$ becomes finite   and  the action is not invariant under local temporal gauge transformations. 
		However, below  the characteristic  energy gap   between Gauss law sectors, we  expect the low-energy physics of  both actions to be identical.  Furthermore,  we see that the bosonic fields have acquired dynamics. 
	\end{itemize}

	\section{Partition function and  absence of sign problem}
	\subsection{Partition function}
	For  the  quantum Monte Carlo  implementation of  our model  we  choose  the  representation:
	\begin{equation}
		\hat{U}_b = e^{i \hat{A}_b}, \qquad \left[ \hat{A}_b,  \hat{E}_{b'}\right] = i \delta_{b,b'}.
	\end{equation}
	and it  is convenient to work in a representation where $\hat A_b$ is diagonal. Omitting the bond index for notational simplicity, we write $\hat A_b |\phi\rangle = \phi |\phi\rangle$ with $\phi \in [0,2\pi)$, while in this representation $\hat E_b = \frac{\partial}{\partial i A_b}$ has eigenstates $|E\rangle$ defined via $\langle \phi | E \rangle = e^{i \phi E}$ with $E \in \mathbb{Z}$. The corresponding resolutions of the identity read
	\begin{equation}
		\int_0^{2\pi} d\phi\, |\phi\rangle\langle\phi| = \frac{1}{2\pi} \sum_{E \in \mathbb{Z}} |E\rangle\langle E| = \hat 1 .
	\end{equation}
	To formulate the path integral, we need the matrix element $\langle \phi' | e^{-\Delta \tau h \hat E_b^2 } | \phi \rangle$. Inserting the resolution of the identity in the $|E\rangle$ basis and using the Poisson summation formula, we obtain the Villain approximation~\cite{villain1975theory,janke1986good}
	\begin{equation}
		\langle \phi' | e^{-\Delta \tau h \hat E_b^2 } | \phi \rangle \propto \exp\left[ -\frac{1}{2 \Delta \tau h} \left( 1 - \cos(\phi - \phi') \right) \right]. \label{eq:eqS3}
	\end{equation}
	Using the Eq.~\eqref{eq:eqS3}, and a similar representation for $\hat{e}_r$ and $\hat{z}_r$, the whole partition is
	\begin{equation}
		\begin{aligned}
			Z=&\int \left(\prod_{\tau_l=1}^{L_\tau} \prod_{x=\{r,b\}} \mathrm{d}\phi_b(\tau_l) \mathrm{d}\varphi_r(\tau_l) \right) \left(\prod_{\tau_l}^{L_\tau} \prod_{x=\{r,b\}} e^{-\frac{1} {2 \Delta \tau h}\left[1-\cos \left(\phi_b(\tau_l)-\phi_b(\tau_l+1)\right)\right]}e^{-\frac{1} {2 \Delta \tau h_m}\left[1-\cos \left(\varphi_r(\tau_l)-\varphi_r(\tau_l+1)\right)\right]}\right)  \\
			& \prod_{\sigma} \mathrm{Tr}_F\prod_{\tau_l=1}^{L_\tau}\left[ \left(\prod_{<i,j>} e^{-\Delta \tau  \hat{c}^\dagger_{i,\sigma} t_{ij} \hat{c}_{j,\sigma}} \right) \left(\prod_{r}e^{-\Delta \tau J \hat{f}^\dagger_{r,\sigma} e^{i \phi_b(\tau_l) } \hat{f}_{r+a_x,\sigma} + h.c.}e^{-\Delta \tau V \hat{c}^\dagger_{r,\sigma} e^{i \varphi_r(\tau_l)  } \hat{f}_{r,\sigma}  + h.c.}\right)\right]\\
			=&\int \left(\prod_{\tau_l=1}^{L_\tau} \prod_{x=\{r,b\}} \mathrm{d}\phi_b(\tau_l) \mathrm{d}\varphi_r(\tau_l) \right) \left(\prod_{\tau_l}^{L_\tau} \prod_{x=\{r,b\}} e^{-\frac{1} {2 \Delta \tau h}\left[1-\cos \left(\phi_b(\tau_l)-\phi_b(\tau_l+1)\right)\right]}e^{-\frac{1} {2 \Delta \tau h_m}\left[1-\cos \left(\varphi_r(\tau_l)-\varphi_r(\tau_l+1)\right)\right]}\right)  \\
			& \prod_{\sigma} \det\left\{\mathrm{I}+B^{\sigma}_{L_\tau}B^{\sigma}_{L_\tau-1}\cdots B^{\sigma}_{2}B^{\sigma}_{1}\right\} \\
			=& \int {\cal D}[\phi,\varphi]\,
			e^{-S_{\rm B}[\phi,\varphi]}
			\prod_{\sigma=\uparrow,\downarrow}
			\det\left\{\mathrm{I}+B^{\sigma}_{L_\tau}B^{\sigma}_{L_\tau-1}\cdots B^{\sigma}_{2}B^{\sigma}_{1}\right\} .
		\end{aligned}  
	\end{equation}
	where $S_{\rm B}[\phi,\varphi]$ is the purely bosonic action of the U(1) fields, and $B^{\sigma}(\tau_l)$ corresponding to $\left(\prod_{\langle i,j\rangle} e^{-\Delta \tau  \hat{c}^\dagger_{i,\sigma} t_{ij} \hat{c}_{j,\sigma}} \right)\left(\prod_{r}e^{-\Delta \tau J \hat{f}^\dagger_{r,\sigma} e^{i \phi_b(\tau_l) } \hat{f}_{r+a_x,\sigma} + \text{h.c.}}e^{-\Delta \tau V \hat{c}^\dagger_{r,\sigma} e^{i \varphi_r(\tau_l) } \hat{f}_{r,\sigma}  + \text{h.c.}}\right)$, whose matrix elements are numbers (functions of the U(1) fields) in the single-particle basis and contain no fermionic operators.

	\subsection{The absence of sign problem}
	
	Then we provide an explicit proof that the determinant weight in our DQMC simulation is non-negative for every configuration of the U(1) fields.
	
	The absence of a sign problem follows from a particle-hole symmetry. We could perform the unitary particle–hole transformation $P$ only on spin-down part:
	
	\begin{equation}
		\begin{aligned}
			\hat{c}_{i,\downarrow} \rightarrow (-1)^i \hat{c}^\dagger_{i,\downarrow} & \qquad \hat{c}^\dagger_{i,\downarrow} \rightarrow (-1)^i \hat{c}_{i,\downarrow} \\
			\hat{f}_{r,\downarrow} \rightarrow (-1)^{r+1} \hat{f}^\dagger_{r,\downarrow} & \qquad \hat{f}^\dagger_{r,\downarrow} \rightarrow (-1)^{r+1} \hat{f}_{r,\downarrow}
		\end{aligned}
	\end{equation}
	After the unitary particle–hole transformation introduced above, which does not change the determinants, the $B$ matrix in the spin-down sector satisfies $P B_l^{\downarrow} P^{-1} = (B_l^{\uparrow})^*$ for every
	configuration, so that the corresponding fermionic determinants are complex conjugates of each other,
	\begin{equation}
		\det\!\left\{\mathrm{I}+B^{\downarrow}_{L_\tau}B^{\downarrow}_{L_\tau-1}\cdots B^{\downarrow}_{2}B^{\downarrow}_{1}\right\} =P \det\!\left\{\mathrm{I}+B^{\downarrow}_{L_\tau}B^{\downarrow}_{L_\tau-1}\cdots B^{\downarrow}_{2}B^{\downarrow}_{1}\right\} P^{-1}
		=\left[\det\!\left\{\mathrm{I}+B^{\uparrow}_{L_\tau}B^{\uparrow}_{L_\tau-1}\cdots B^{\uparrow}_{2}B^{\uparrow}_{1}\right\}\right]^*,
	\end{equation}
	and the weight is a real positive number: $\prod_{\sigma=\uparrow,\downarrow}
	\det\left\{\mathrm{I}+B^{\sigma}_{L_\tau}B^{\sigma}_{L_\tau-1}\cdots B^{\sigma}_{2}B^{\sigma}_{1}\right\}=|\det\{\mathrm{I}+B^{\uparrow}_{L_\tau}\cdots B^{\uparrow}_{1}\}|^2 \ge 0$. 
	
	Since the bosonic weight $e^{-S_{\rm B}[\phi,\varphi]}$ is also real and
	positive, the full Monte Carlo weight is non-negative for every field
	configuration. This establishes the absence of a fermion sign problem~\cite{wu2005sufficient}.
	
	\section{Gauge-invariant periodic-Anderson benchmark and Kubo formula}
		\subsection{Noninteracting multiband Hamiltonian}
		
		We consider the gauge-invariant weak-fluctuation limit discussed in
		the End Matter. In the translationally invariant zero-flux sector
		$W_{r,r+\hat{x}}=1$, the model reduces to an $(L+1)$-band periodic
		Anderson Hamiltonian for the conduction electrons and the
		gauge-neutral composite fermion
		$\Psi_r=z_r^\dagger f_r$:
		\begin{equation}
			\begin{aligned}
				\hat{H}_0
				={}&
				\sum_{i,j}
				\left(
				t_{ij}\hat{c}_i^\dagger\hat{c}_j
				+\text{h.c.}
				\right)
				+
				J\sum_r
				\left(
				\hat{\Psi}_r^\dagger
				\hat{\Psi}_{r+\hat{x}}
				+\text{h.c.}
				\right)
				+
				V\sum_r
				\left(
				\hat{c}_r^\dagger\hat{\Psi}_r
				+\text{h.c.}
				\right)
				\\
				={}&
				\sum_k\sum_{m,n=1}^{L+1}
				\hat{d}_{k,m}^\dagger
				H_{mn}(k)
				\hat{d}_{k,n}
				\\
				={}&
				\sum_k\sum_{n=1}^{L+1}
				E_n(k)
				\hat{\gamma}_{k,n}^\dagger
				\hat{\gamma}_{k,n}.
			\end{aligned}
	\end{equation}
	Here we have defined
	\begin{equation}
		\hat{d}_{k,n}^{\dagger}=
			\begin{cases}
				\hat{c}_{k,n}^{\dagger}, & n=1,\dots,L,\\[2pt]
				\hat{\Psi}_{k}^{\dagger}, & n=L+1.
		\end{cases}
	\end{equation}
	and diagonalized the single-particle Hamiltonian as
	$H(k)=U(k) E(k) U^{\dagger}(k)$, where
	$E(k)=\mathrm{diag}\{E_{1}(k),\dots,E_{L+1}(k)\}$ and
	$\gamma_{k}^{\dagger}=d_{k}^{\dagger}U(k)$.
	The equilibrium single-particle density matrix in this basis is
	\begin{equation}
		G_{rs}(k)
		=\langle d_{k,r}^{\dagger} d_{k,s}\rangle
		=\sum_{n} U_{r n}(k)\,f\!\bigl(E_{n}(k)\bigr)\,
		U^{\dagger}_{n s}(k),
		\label{eq:Gks_def}
	\end{equation}
	where $f(E)=1/(e^{\beta E}+1)$ is the Fermi function.

	\begin{figure}[htbp]
		\centering
		\includegraphics[width=0.7\linewidth]{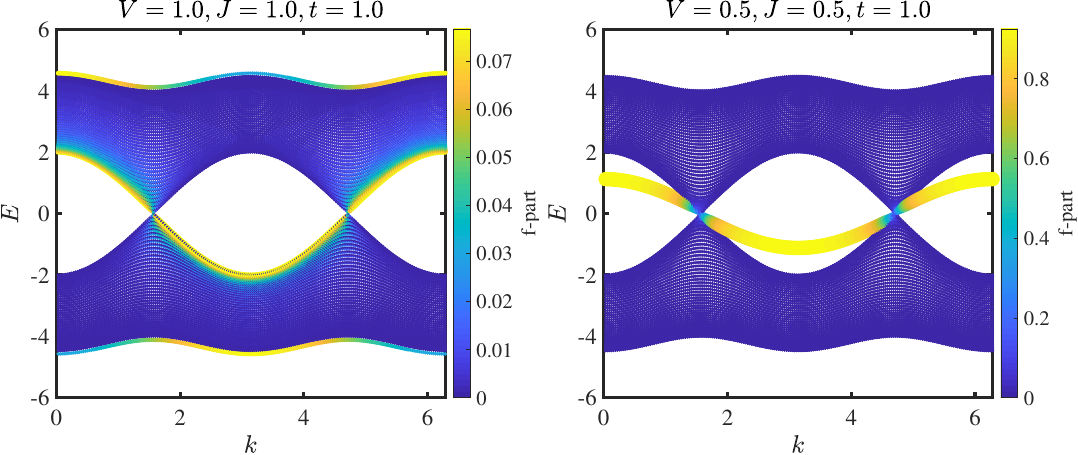}
		\caption{
			Band structure of the one-dimensional tight-binding model for two sets of parameters.
			The dispersions are obtained by diagonalizing the Bloch Hamiltonian, and the color scale encodes the $\Psi$-electron component  of the $i$-th band at momentum $k$, quantified by $|U_{L+1,i}(k)|^{2}$; brighter (yellow) points correspond to states with larger $f$-electron weight .
			Left: $V = 1.0$, $J = 1.0$, $t = 1.0$.
			Right: $V = 0.5$, $J = 0.5$, $t = 1.0$.
		}
		\label{fig:figs2}
	\end{figure}
	
	It is convenient to introduce projection operators onto the
	single-particle eigenstates,
	\begin{equation}
		P_n(k) = U(k)_{:,n}\,U^{\dagger}(k)_{n,:},
		\qquad
		H(k) = \sum_{n} E_{n}(k)\,P_{n}(k),
	\end{equation}
	so that
	\begin{equation}
		G(k)=\sum_{n} f\!\bigl(E_{n}(k)\bigr)\,P_{n}(k),
		\qquad
		e^{-i t H(k)}=\sum_{n} e^{-i t E_n(k)}\,P_n(k).
	\end{equation}
	
	For this tight-binding model, the band dispersion is obtained by diagonalizing the Bloch Hamiltonian, and the color scale represents the $f$-electron weight of each eigenstate, $|U_{L+1,i}(k)|^{2}$.
	
	In the main text we focus on the bare parameters $V=1$ and $J=1$. When the parameter $h$ is finite but not too large, gauge fluctuations suppress the relevant mean-field amplitudes, leading to renormalized couplings $V_{\mathrm{eff}}<V$ and $J_{\mathrm{eff}}<J$. To illustrate the corresponding band-structure evolution, Fig.~\ref{fig:figs2} compares a representative choice $(V,J)=(1.0,1.0)$ with a reduced set $(V,J)=(0.5,0.5)$ (both at $t=1.0$). The reduced-parameter case closely mimics the regime discussed in the first panel of Fig.~2(a) in the main text: a pronounced hybridized band emerges, with substantial $f$-electron character over a broad range of momenta, reflecting strong $c$--$f$ mixing.
	
	\subsection{Current operator and kinetic energy}
	
	We couple the $f$-fermion hopping to an external vector potential
	$\mathbf{A}(t)$ via the Peierls substitution
	\begin{equation}
		H
		=\sum_{i,j} \bigl( t_{ij}\,c_i^{\dagger} c_j + \text{h.c.} \bigr)
		+ J\sum_{\langle r,r'\rangle}
		\Bigl( \Psi_{r}^{\dagger} \Psi_{r'}
		e^{\tfrac{2\pi i}{\Phi_0}\int_{r}^{r'} \mathbf{A}(t)\cdot d\mathbf{\ell}}
		+ \text{h.c.} \Bigr)
		+ V\sum_{r} \bigl( c_r^{\dagger} \Psi_r + \text{h.c.} \bigr),
	\end{equation}
	where $\Phi_0$ is the flux quantum and we have in mind
	nearest-neighbor hopping along the $x$ direction.
	
	Expanding to linear order in $\mathbf{A}(t)$ we obtain
	the interaction Hamiltonian
	\begin{equation}
		H_{I}(t)
		=\frac{2\pi}{\Phi_0}\,
		\hat{\mathbf{J}}_{p}\cdot\mathbf{A}(t),
	\end{equation}
	which defines the paramagnetic current operator
	\begin{equation}
		\hat{\mathbf{J}}_{p}
		= i \sum_{i,j} (\mathbf{r}_j-\mathbf{r}_i)\,
		J_{ij}\,\Psi_i^{\dagger} \Psi_j.
	\end{equation}
	The diamagnetic contribution is
	\begin{equation}
		\mathbf{J}_{D}(t)
		= -\sum_{i,j} (\mathbf{r}_j-\mathbf{r}_i)\,
		J_{ij}\,\Psi_i^{\dagger} \Psi_j\,
		[(\mathbf{r}_j-\mathbf{r}_i)\cdot\mathbf{A}(t)],
	\end{equation}
	and the current carried by the composite-fermion channel (per lattice length $L$) reads
	\begin{equation}
		\mathbf{J}^{\mathrm{tot}}(t)
		= -\frac{2\pi}{\Phi_0 L}
		\left[ \hat{\mathbf{J}}_{p}
		+\frac{2\pi}{\Phi_0}\,\mathbf{J}_{D}(t) \right].
	\end{equation}
	In what follows we focus on the longitudinal component and drop
	vector notation for brevity.
	
	For nearest-neighbor hopping along the chain, the paramagnetic
	current can be written in momentum space as
	\begin{equation}
		\hat{J}_{p}
		= iJ\sum_{i}
		\bigl( \Psi_i^{\dagger} \Psi_{i+\hat{x}} - \Psi_{i+\hat{x}}^{\dagger} \Psi_{i} \bigr)
		= -2J \sum_{k} \sin k\, \Psi_{k}^{\dagger} \Psi_{k}
		\equiv \sum_{k,n,m} B_{nm}(k)\,d_{k,n}^{\dagger} d_{k,m},
	\end{equation}
	with
	\begin{equation}
		B_{nm}(k) = -2J \sin k\; \delta_{n,L+1}\delta_{m,L+1}.
	\end{equation}
	Similarly, the kinetic-energy term is
	\begin{equation}
		K
		=\sum_{i,j} (\mathbf{r}_j-\mathbf{r}_i)^{2}
		J_{ij}\,\Psi_i^{\dagger} \Psi_j
		=\sum_{k} 2J\cos k\, \Psi_{k}^{\dagger}\Psi_{k}
		\equiv \sum_{k,n,m} C_{nm}(k)\,d_{k,n}^{\dagger} d_{k,m},
	\end{equation}
	with
	\begin{equation}
		C_{nm}(k) = 2J \cos k\; \delta_{n,L+1}\delta_{m,L+1}.
	\end{equation}
	Using Eq.~\eqref{eq:Gks_def}, its expectation value is
	\begin{equation}
		\langle K \rangle
		= \sum_{k} \mathrm{Tr}\bigl[ C(k)\,G(k) \bigr].
	\end{equation}
	
	\subsection{Current--current correlation function}
	
	We set $\hbar=1$ and define the retarded current--current
	correlation function
	\begin{equation}
		\Lambda(\omega)
		= i\int_{0}^{\infty} dt\,
		e^{i(\omega+i\eta)t}\,
		\bigl\langle [\hat{J}_{p},\hat{J}_{p}^{H}(-t)] \bigr\rangle_{0},
	\end{equation}
	where the Heisenberg operator is evolved with $H_0$ and
	$\langle\cdots\rangle_{0}$ denotes the thermal average.
	Writing $\hat{J}_{p}$ in the $d$-basis and using standard
	Wick contractions, one obtains
	\begin{equation}
		\begin{aligned}
			\Lambda(\omega)
			&= i\int_{0}^{\infty} dt\,
			e^{i(\omega+i\eta)t}
			\sum_{k}
			\mathrm{Tr}\Bigl\{
			G(k)\,
			\bigl[
			B(k),
			e^{-it H(k)} B(k) e^{it H(k)}
			\bigr]
			\Bigr\} \\
			&= i\sum_{k}\sum_{m,n}
			\int_{0}^{\infty} dt\,
			e^{i(\omega+i\eta+E_{m}(k)-E_{n}(k))t}\,
			\mathrm{Tr}\Bigl\{
			G(k)\,
			\bigl[B(k),P_{n}(k) B(k) P_{m}(k)\bigr]
			\Bigr\}.
		\end{aligned}
	\end{equation}
	Evaluating the time integral and using
	$G(k)=\sum_{x} f(E_{x}(k))P_{x}(k)$ yields
	\begin{equation}
		\Lambda(\omega)
		= \sum_{k}\sum_{m,n}
		\frac{f(E_{n}(k))-f(E_{m}(k))}
		{\omega+i\eta+E_{m}(k)-E_{n}(k)}\,
		\mathrm{Tr}\Bigl[
		B(k) P_{n}(k) B(k) P_{m}(k)
		\Bigr].
		\label{eq:Lambda_final}
	\end{equation}
	
	\subsection{Optical conductivity}
	
	Finally, the longitudinal optical conductivity follows from
	the Kubo formula
	\begin{equation}
		\sigma(\omega)
		= \left( \frac{2\pi}{\Phi_0} \right)^{2}
		\frac{1}{L}\,
		\frac{1}{i(\omega+i\eta)}
		\Bigl[\,\langle K \rangle + \Lambda(\omega)\Bigr],
	\end{equation}
	with $\langle K\rangle$ and $\Lambda(\omega)$ given above.
	Equivalently, separating real and imaginary parts,
	\begin{equation}
		\begin{aligned}
			\sigma(\omega)
			&= \left( \frac{2\pi}{\Phi_0} \right)^{2}
			\frac{1}{L}
			\frac{1}{i(\omega+i\eta)}
			\Biggl[
			\sum_{k,n} f\!\bigl(E_{n}(k)\bigr)
			\mathrm{Tr}\bigl[C(k) P_{n}(k)\bigr]
			+ \sum_{k,m,n}
			\frac{f(E_{n}(k))-f(E_{m}(k))}
			{\omega+i\eta+E_{m}(k)-E_{n}(k)} \\
			&\hspace{5cm}\times
			\mathrm{Tr}\bigl[ B(k) P_{n}(k) B(k) P_{m}(k) \bigr]
			\Biggr]
		\end{aligned}
	\end{equation}
	where in the last line we have used
	Eq.~\eqref{eq:Lambda_final} together with
	$B_{nm}(k)=-2J\sin k\,\delta_{n,L+1}\delta_{m,L+1}$ and
	$C_{nm}(k)=2J\cos k\,\delta_{n,L+1}\delta_{m,L+1}$.
	
	\begin{figure}[htbp]
		\centering
		\includegraphics[width=0.5\linewidth]{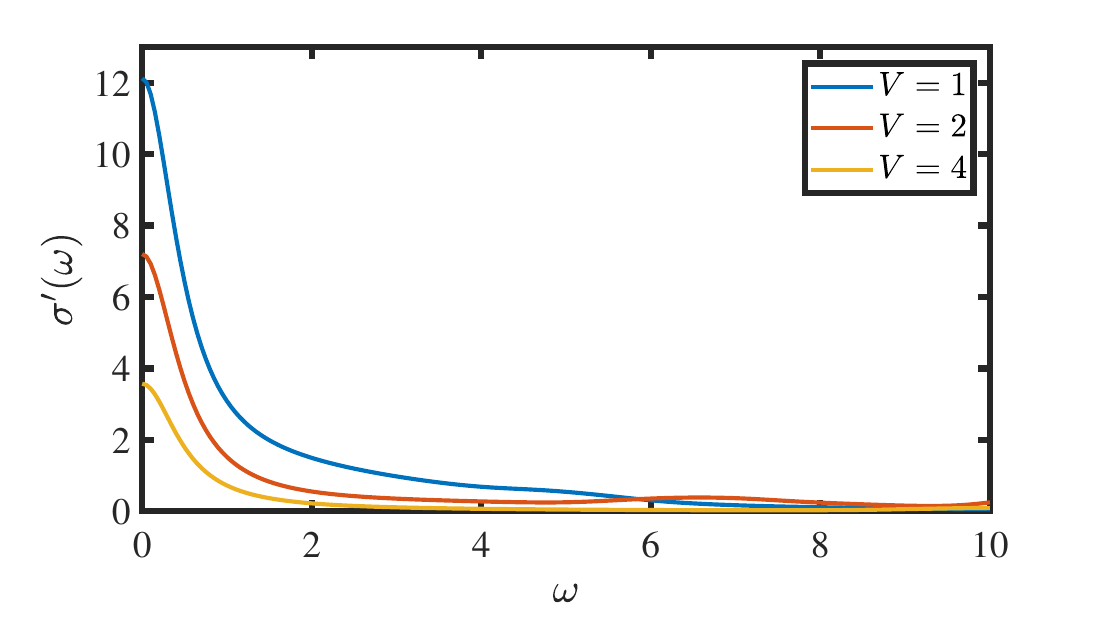}
		\caption{For $L=100$, $J=0.5$, $t=1$, $\beta=L$, $\Delta \omega =0.05$, and $\eta = 10\Delta \omega$, the mean-field results are shown in the figure. It can be seen that when $V$ takes moderate values, the zero-frequency conductivity remains finite, corresponding to the Kondo phase. In contrast, when $V$ becomes too large, the hybridization band becomes overly flat, leading to a suppression of the conductivity.
		}
		\label{fig:s3}
	\end{figure}	
	\newpage
	\section{Additional data}
	
	In the main text, we presented the low-temperature optical conductivity $\sigma'(\omega)$ of the composite-fermion chain, the composite-fermion spectral function $A_{\Psi}(k,\omega)$, and the dynamical spin structure factor of the $f$ chain $S_f(q,\omega)$ for three representative values of the transverse field $h$ (see Figs.~2--4
	). In this Supplementary Material we extend these results in two complementary directions.
	First, we provide a denser scan of $h$ at fixed low temperature (the same $\beta=5$ as used in the main text) in order to better resolve how the fermionic, spin, and transport responses evolve across the transition from the Kondo-coherent regime to the Kondo-breakdown regime. 
	
	Second, we select three representative values of $h$ (one in the Kondo-coherent regime, one near the transition, and one in the Kondo-breakdown regime) and present results at multiple temperatures. This temperature-dependent dataset clarifies how the Drude response, hybridization features in $A_{\Psi}(k,\omega)$, and the spin dynamics are thermally broadened and redistributed.
	
	Third, we present an additional finite-size analysis of the composite-fermion spectral weight near the Kondo-breakdown transition. As shown in Fig.~\ref{fig:s7}, finite-size effects are indeed visible at the lowest temperatures but remain relatively weak. The data support the interpretation proposed in the manuscript that the suppression of the low-temperature composite-fermion spectral weight within the Kondo phase is an intrinsic crossover rather than a finite-size effect. In particular, when the temperature exceeds the characteristic gauge-fluctuation scale set by $h$, the gauge field is effectively frozen on thermal time scales and the system is well approximated by the non-interacting periodic Anderson model, resulting in a coherent composite-fermion band with a large density of states. Upon lowering the temperature below this scale, dynamical gauge-field fluctuations become increasingly important and progressively suppress the low-energy composite-fermion spectral weight while preserving Kondo coherence. 		
	In the stochastic analytic continuation, we sample
		$A_C(\omega)=\omega\coth(\beta\omega/2)\sigma'(\omega)$.
		Its normalization is fixed by the exact sum rule
		$\int_0^\infty d\omega\,A_C(\omega)=\pi\Lambda(0)$.
		The optical conductivity is then obtained as
		$\sigma'(\omega)=\tanh(\beta\omega/2)A_C(\omega)/\omega$.
		Thus, the sum rule is included directly in the continuation.

	\subsection{Extended low-$T$ scan in $h$.}
	Figure~\ref{fig:s4} shows $\sigma'(\omega)$ at fixed $\beta$ for an extended set of $h$ values. In the small-$h$ regime the spectrum is dominated by a Drude-like response near $\omega\simeq 0$, consistent with metallic transport of itinerant composite quasiparticles in the Kondo-coherent regime. As $h$ increases, the low-frequency weight is progressively suppressed and spectral weight is transferred to finite frequencies, signaling the gradual loss of coherent transport and the onset of Mott-insulating behavior in the $f$ sector. The corresponding evolution in the composite-fermion spectrum and spin spectrum is also shown in Fig.~\ref{fig:s4}. In particular, the coherent hybridized band in $A_{\Psi}(k,\omega)$ weakens with increasing $h$ and eventually gives way to two separated incoherent bands, while the spin response evolves toward that of an effective one-dimensional Heisenberg chain, corroborating the Kondo-breakdown interpretation discussed in the main text.

	\begin{figure}[!htbp]
		\centering
		\includegraphics[width=\linewidth]{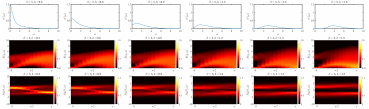}
		\caption{\textbf{Low-temperature $h$ scan.}
			Low-temperature cuts (fixed inverse temperature $\beta=5$) of three key observables as a function of $h$:
			(top row) the real part of the optical conductivity of the composite-fermion chain, $\sigma'(\omega)\equiv \mathrm{Re}\,\sigma(\omega)$;
			(middle row) the dynamical spin structure factor of the $f$ chain, $S_f(q,\omega)$;
			(bottom row) the composite-fermion spectral function, $A_{\Psi}(k,\omega)$.
			Each column corresponds to a different value of $h$ (from left to right: increasing $h$), spanning the evolution from the Kondo-coherent regime (small $h$) to the Kondo-breakdown regime (large $h$).
			In the small-$h$ regime, $\sigma'(\omega)$ exhibits a pronounced low-frequency (Drude-like) response and $A_{\Psi}(k,\omega)$ shows a coherent hybridized band, while at larger $h$ the low-frequency optical weight is strongly suppressed, the hybridization features in $A_{\Psi}(k,\omega)$ disappear, and $S_f(q,\omega)$ evolves toward a well-defined dispersive mode consistent with local-moment dynamics.}
		
	\end{figure}
	
	\newpage
	
	\subsection{Temperature dependence at representative $h$.}
	To complement the low-$T$ $h$ scan, we fix three representative values of $h$ and study the temperature evolution of $\sigma'(\omega)$, $A_{\Psi}(k,\omega)$, and $S_f(q,\omega)$ (Figs.~\ref{fig:s4}--\ref{fig:s6}). These supplementary data provide a more complete view of how gauge fluctuations controlled by $h$ and thermal fluctuations cooperate to suppress Kondo hybridization and drive orbital-selective Mott localization in the $f$ sector.

	\begin{figure}[htbp]
		\centering
		
		\includegraphics[width=\linewidth]{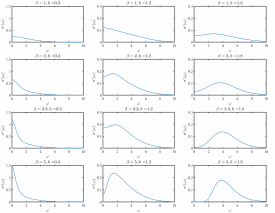}
		
		\caption{\textbf{Temperature evolution of the optical conductivity at representative $h$.}
			$\sigma'(\omega)\equiv \mathrm{Re}\,\sigma(\omega)$ at three representative values $h=0.3,1.2,1.8$ for a range of inverse temperatures $\beta=1,2,3.3,5$.}
		
		\label{fig:s4}
	\end{figure}

	\begin{figure}[htbp]
		\centering
		\includegraphics[width=\linewidth]{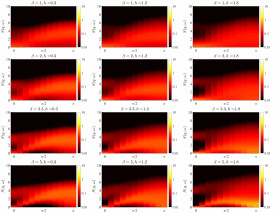}

		\caption{\textbf{Temperature evolution of the spin dynamics at representative $h$.}
			Dynamical spin structure factor $S_f(q,\omega)$ at different $h$ for the same set of temperatures as in Fig.~\ref{fig:s4}.}
		
		\label{fig:s5}
	\end{figure}
	
	\begin{figure}[htbp]
		\centering
		\includegraphics[width=\linewidth]{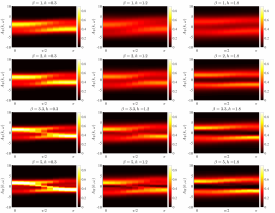}
		\caption{\textbf{Temperature evolution of the composite-fermion spectrum at representative $h$.}
			Composite-fermion spectral function $A_{\Psi}(k,\omega)$ at different $h$ for the same set of temperatures as in Fig.~\ref{fig:s4}.}
		
		\label{fig:s6}
	\end{figure}
	\FloatBarrier
	
	\subsection{Finite-size analysis of the composite-fermion spectral weight}
	\begin{figure}[!htbp]
		\centering
		\includegraphics[width=0.8\linewidth]{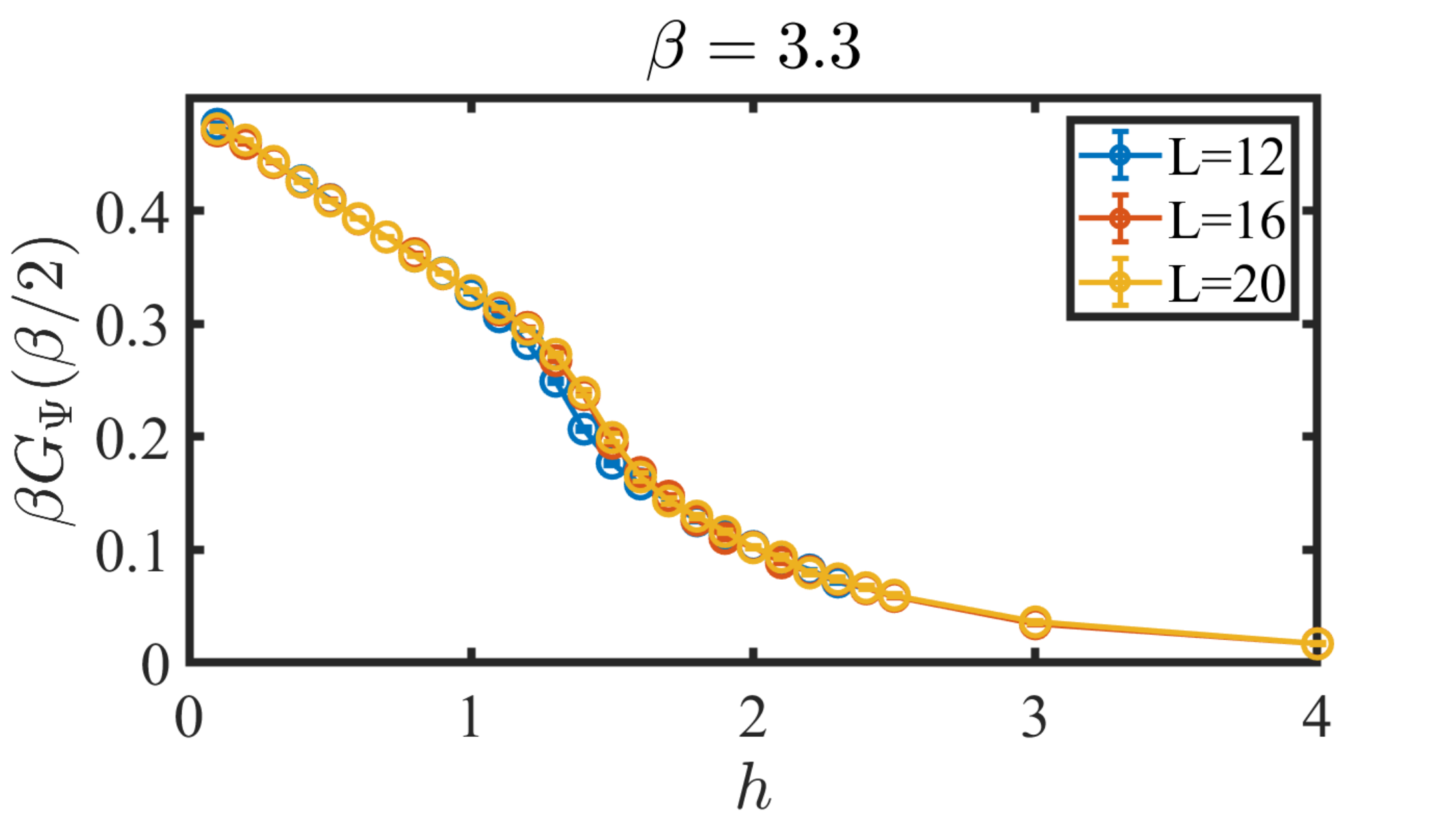}
		\caption{\textbf{Finite-size effects}
			Finite-size dependence of the composite-fermion spectral weight $N_{\Psi}(\omega=0)\simeq\beta G_{\Psi}(\beta/2)$ for several system sizes near the Kondo-breakdown transition. Although finite-size effects become visible at the lowest temperatures, their magnitude remains relatively modest. Combined with the analysis presented in the main text, these results support the interpretation that the low-temperature suppression of the composite-fermion spectral weight within the Kondo phase reflects the onset of gauge-field fluctuations below the characteristic energy scale set by $h$, rather than a finite-size artifact.
		}
		\label{fig:s7}
	\end{figure}
	
	\clearpage
	
\end{widetext}

\end{document}